\documentclass[a4paper, 12pt]{article}
\usepackage{moreverb}
\usepackage{url}
\usepackage{grffile}
\usepackage{lineno}
\usepackage{lipsum}
\usepackage[UKenglish]{isodate}
\usepackage{bm} 
\usepackage{natbib}
\usepackage{amssymb}
\usepackage{graphicx}
\usepackage{xcolor}

\newcommand\BibTeX{{\rmfamily B\kern-.05em \textsc{i\kern-.025em b}\kern-.08em
T\kern-.1667em\lower.7ex\hbox{E}\kern-.125emX}}
\newcommand\MiKTeX{{\rmfamily M\kern-.05em \textsc{i\kern-.025em K}\kern-.08em
T\kern-.1667em\lower.7ex\hbox{E}\kern-.125emX}}
\newcommand\PracTeX{{\rmfamily P\kern-.05em \textsc{r\kern-.025em a\kern-.025em
c}\kern-.08em
T\kern-.1667em\lower.7ex\hbox{E}\kern-.125emX}}

\newcommand{\gs}{{j=1,\allowbreak  2, \allowbreak \ldots, \allowbreak G}}
\newcommand{\is}{{i=1,\allowbreak  2, \allowbreak \ldots, \allowbreak n}}
\newcommand{\lmax}{\lambda_{\max}}
\newcommand{\lmin}{\lambda_{\min}}
\newcommand{\bx}{\mbox{$\boldsymbol{x}$}}
\newcommand{\bX}{\mbox{$\boldsymbol{X}$}}

\newcommand{\var}{\rm{var}}
\newcommand{\R}{\mathbb{R}}

\def\argmin{\mathop{\rm arg\,min}\limits}

\begin{document}
\title{An adequacy approach for deciding the number of clusters for OTRIMLE 
robust Gaussian mixture based clustering}
\author{Christian Hennig\thanks{christian.hennig@unibo.it, Dipartimento di Scienze Statistiche, Universita di Bologna, Italy} \and Pietro Coretto\thanks{Department of Economics and Statistics
University of Salerno, Italy}}
\maketitle

\begin{abstract}
We introduce a new approach to deciding the number of clusters. The approach
is applied to Optimally Tuned Robust Improper Maximum Likelihood Estimation
(OTRIMLE; \cite{CorHen16}) of a Gaussian mixture model allowing for observations to be 
classified as ``noise'', but it can be applied to other clustering methods as
well. The quality of a clustering is assessed by a statistic $Q$ that measures
how close the within-cluster distributions are to elliptical unimodal 
distributions that have the only mode in the mean. This nonparametric measure 
allows for non-Gaussian clusters as long as they have a good quality
according to $Q$. The simplicity of a model is
assessed by a measure $S$ that prefers a smaller number of clusters unless 
additional
clusters can reduce the estimated noise proportion substantially. The simplest
model is then chosen that is adequate for the data in the sense that its 
observed value of $Q$ is not significantly larger than what is expected for data
truly generated from the fitted model, as can be assessed by parametric 
bootstrap. The approach is compared with model-based clustering using the 
Bayesian Information Criterion (BIC) and the Integrated Complete 
Likelihood (ICL) in a simulation study and on two datasets
of scientific interest.~
\end{abstract}


\section{Introduction}
\label{sec:intro}
We introduce an approach for finding a suitable number of clusters for use 
with Optimally Tuned Robust Improper Maximum Likelihood (OTRIMLE)
clustering \citep{CorHen16,CorHen17}, which attempts to find approximately 
Gaussian distributed clusters allowing for some observations to be classified
as noise or outliers. The approach in its general form is very flexible and
can be adapted to other clustering methods and other types of clusters, but we
focus on its use with OTRIMLE here. {\color{black}  The approach is based on
adequacy testing of a fitted model by using parametric bootstrap. An early 
forerunner of this approach is \cite{McLachlan87}.}

A key issue with choosing the number of clusters is that 
model assumptions never 
hold precisely in reality. It is therefore 
important that statistical methods produce reasonable
results even if the model assumptions are violated. The problem with this is 
that what the 
method tries to estimate is usually defined in terms of the nominal (assumed) 
model, and if the model does not hold, it is not always 
clear what a ``reasonable'' result would be. 
If clusters are supposed to be (approximately) Gaussian,
using a Gaussian mixture model for clustering \citep{BanRaf93} looks 
attractive. Estimation of the number of clusters for this is often done using
the Bayesian Information Criterion (BIC), e.g., in the R package mclust
\citep{mclust16}. The BIC has been 
proven to be consistent for estimating the number of mixture components
\citep{Keribin00} under some rather restrictive assumptions, and is believed to
be more generally consistent. {\color{black}  This may be seen as an advantage,} 
but is in fact a 
problem {\color{black}  if the aim is to interpret the mixture components as 
clusters rather than just finding a good approximating Gaussian mixture for
the data}. A Gaussian mixture model with a sufficiently large number of mixture 
components can approximate almost any distribution arbitrarily well 
(for a recent precise version of this statement and a discussion of 
some older versions see \cite{NNCM20}), and in reality, clusters are not 
precisely Gaussian. This means that if the number of 
observations $n$ becomes larger, a consistent method for estimating the number
of mixture components can be expected to add 
mixture components in order to fit the real distribution 
better, and ultimately several components will fit 
an approximately but not precisely Gaussian subset of the data that intuitively
would qualify as a single cluster, in turn overestimating the number of 
clusters. {\color{black} For the same reason, 
a likelihood ratio test will reject a single
Gaussian distribution for such clusters with large probability for $n$ large 
enough.}
This has also been observed in practice for the BIC
\citep{Hennig10}. The estimation of the number of clusters is therefore 
affected by 
violations of the model assumptions in a more critical way than most standard
statistical estimation problems.

{\color{black} The precise definition of outliers/noise in cluster analysis
is another issue.}  
There is an ambiguity between noise and clusters in two respects.
Firstly, it is not clear how large a group of outliers has to be in order to be 
interpreted as a cluster on its own, and secondly, there may be very widely 
spread observations that can be well approximated by a Gaussian distribution 
with a very low density everywhere but may more appropriately be 
interpreted as noise than as a cluster, depending on the subject matter and the
meaning of the data. 
{\color{black} Not allowing for noise classification does not solve these 
issues as long as there are observations that should appropriately be
interpreted as outliers; integrating them into regular 
clusters affects the estimation of these clusters and 
can also lead to misclassification of other observations.} 

A further issue is that to some extent more mixture components can be traded 
off against more flexible covariance matrices. Too flexible covariance matrices
are already an issue for a fixed number of mixture components because of 
potential degeneracy or near-degeneracy of the likelihood \citep{GEGGIMI18}. 
 
The consequence is that finding an appropriate number of clusters should not be
seen as a well-defined estimation problem in a statistical model. Rather it 
essentially requires decisions by the user: how much better approximation of the
data, how much simpler covariance matrix structure that is 
less prone to degeneracy, 
and what decrease of the noise proportion, would justify adding another 
mixture component? A method that does not require any user input such as the
BIC should not be trusted naively. These issues are acknowledged for example
by the authors of the R-package \texttt{tclust} for robust trimmed clustering 
\citep{FrGaMa12}, who do not offer an automatic method for choosing the number
of clusters, but rather some graphical displays that allow the used to track
the different aspects to be traded off against each other. 

On the other hand, in many situations users do not have sufficient background
knowledge to make all the required decisions in a well founded manner, and also,
{\color{black} for systematically
evaluating the quality of an approach,
automation that does not require manual adaptation to every data 
set is required.} For this reason we offer an approach that
allows the user to make the required tuning decisions, but we also suggest some
default choices to give the user a starting point and to enable evaluation by
simulation. {\color{black} Ultimately an optimal tuning should depend on 
knowledge about the subject matter background and the aim of clustering.}

The approach is based on the concept of ``adequacy'' introduced by 
\cite{Davies95}. According to this concept, a model (Davies' use of the term
``model'' includes
specific distributions with given parameter values) is adequate for a data set
with respect to a statistic $Q$ if the value of $Q$ on the data set 
is ``typical'' for data sets generated by the model. {\color{black} This basically
means that a significance test based on $Q$ does not
reject the model.} $Q$ is chosen to reflect the sense in which the model needs
to ``fit'' the data in a given application rather than following optimality
considerations such as those by Neyman-Pearson; 
more than one test statistic can 
be chosen and can be combined using Bonferroni's correction. Unless the 
distribution of $Q$ on the model can be handled analytically, parametric 
bootstrap can be applied to approximate this distribution.
The selection of the number of clusters
is a model selection problem, and Davies recommends to select the simplest 
model that is adequate for the data \citep{DavKov01}, 
which could be the model with the lowest
number of mixture components, but see Section \ref{sec:simplicity}. 
Note that whenever
a mixture with a low number of mixture components fits the data adequately, the
data could also be fit by a model with more mixture components (one could just 
add small components around single observations), which
means that the data actually cannot distinguish between a model with a 
small number of well fitting mixture components and a model with a larger 
number of components, despite the fact that automatic rules such as the BIC
{\color{black} may be interpreted by users as if this were possible. 
The simplest model 
that fits can be preferred for reasons of parsimony, avoidance of overfitting,
potentially better 
suitability for generalisation, and ease of interpretation.}

{\color{black} 
The problem of choosing the number of clusters is notoriously difficult and 
has been treated by many authors. In the mixture context, several alternatives 
to the BIC have been considered. One popular alternative is the
Integrated Completed Likelihood (ICL; \cite{BiCeGo00}), which as the method 
introduced here is meant to address the issue that the BIC can choose 
too many mixture components for non-Gaussian clusters. Other alternatives to 
the BIC include the AICmix \citep{HuWaFo15} and MSCAD \citep{CheKha08}, 
testing different model orders against each other by means of bootstrapping the
likelihood ratio
\citep{McLachlan87,FenMcC96} or theoretically \citep{ChLiFu12}, and 
Bayesian
approaches \citep{XieXu20} with more references in \cite{FrCeRo20}. An 
alternative approach to fit non-Gaussian clusters with Gaussian mixtures is 
merging of mixture components \citep{Hennig10,BRCLG10,MaFrGr17}. 
\cite{HenLin15} use parametric bootstrap from a null model for homogeneous data
for choosing the number of clusters. Section 4.3 of \cite{Ritter14} reviews
versions of the BIC for robust clustering with trimming. 

Some other work 
comparing a model fit on data with data generated from
parametric bootstrap comprises \cite{WaRaGoIv04}, and 
posterior predictive checking in a Bayesian framework \citep{Meng94,GeMeSt96}.}

The OTRIMLE method is introduced in Section \ref{sec:otrimle}. Section 
\ref{sec:approach} gives an outline of the approach for deciding the number of 
clusters. This approach requires a number of decisions by the user. 
Section 
\ref{sec:statistic} contains proposals for these decisions. Particularly,
a statistic $Q$ is proposed that measures to what extent 
the found clusters in a data set for a given number of clusters 
qualify as ``adequate''.
In Section 
\ref{sec:experiments}, we compare the method with the BIC {\color{black} and
ICL} for Gaussian mixtures,
Gaussian mixtures with noise, mixtures of t- and skew t-distributions. Section
\ref{sec:conclusion} concludes the paper.

We are very happy to be invited to contribute this paper to a Special Issue in 
honour of Geoff McLachlan, {\color{black} who is a pioneer of the use of
parametric bootstrap for estimating the number of mixture components 
\citep{McLachlan87}, mixtures of t- and skew t-distributions and their use for 
accommodating outliers \citep{PeeMcL00,LeeMcL13}, and who contributed to making
statements about the approximation of arbitrary distributions by mixtures
precise \citep{NNCM20}. He has also contributed to inspiring and
improving work of ours by 
many valuable remarks, for which we are very grateful.}

\section{The OTRIMLE approach to robust clustering} \label{sec:otrimle}
When using mixture models for cluster analysis, usually mixtures of 
families of distributions are considered that formalise the idea of a
homogeneous cluster. Every mixture component is then interpreted as modelling
a cluster, and the number of mixture components corresponds to the number of
clusters (there are exceptions to this, see \cite{Hennig10}). 

The most popular choice for continuous data is the family of Gaussian 
distributions. 
A standard Gaussian mixture model assumes data $\bx_1,\ldots,\bx_n$ to be generated
independently identically distributed from a distribution with density
\begin{equation}\label{eq:gaussmix}
f(\bx; {\bm{\theta}})=\sum_{g=1}^{G} \pi_g \phi_p(\bx; {\bm{\mu}}_g, {\bf \Sigma}_g),
\end{equation}
where $\phi_p(\cdot; {\bm{\mu}}, {\bf \Sigma})$ is the $p$-variate Gaussian density with mean ${\bm{\mu}}$ and covariance matrix ${\bf \Sigma}$, $\pi_g \in [0,1]$ for $\gs$, $\sum\nolimits_{i=1}^G \pi_g = 1$, and ${\bm{\theta}}$ is the parameter vector collecting all 
$\pi_g, {\bm{\mu}}_g, {\bf \Sigma}_g, \gs$. For given $G$, the parameters ${\bm{\theta}}$ can be 
estimated by maximum likelihood (ML). More precisely, a global optimum is often 
not available, and algorithms such as the EM-algorithm are used that find a 
local 
optimum of the likelihood. Given estimators (here denoted $\hat{\bm{\theta}}, \hat\pi_g, 
\hat{\bm{\mu}}_g, \hat{\bf \Sigma}_g, \gs$), probabilities that observations $\bx_i, \is,$ were
generated by mixture component $g$ can be estimated as
\begin{equation}
  \label{eq:postprob}
  \hat p_{ig}=\frac{\hat\pi_g\phi_p(\bx_i; \hat{\bm{\mu}}_g, \hat{\bf \Sigma}_g)}
{f(\bx_i; \hat{\bm{\theta}})},
\end{equation}
and observation $i$ can be assigned to the mixture component $g$ that maximises
$\hat p_{ig}$. This is implemented in the R-package \texttt{mclust} \citep{BanRaf93,mclust16}, along with a number of models defined by various constraints on
the within-component covariance matrices. The \texttt{mclust}-approach 
for deciding the
number of mixture components $G$ and the covariance matrix model is to minimise
the Bayesian Information Criterion (BIC), 
\begin{equation}
  \label{eq:bic}
 \mbox{BIC}=k\ln(n)-2\ln(\hat L_n),
\end{equation}
where $k$ is the number of free parameters ($k=(G-1)+pG+p(p+1)G/2$ for a model
with fully free covariance matrices), and $\hat L_n$ is the maximised likelihood
for the model under investigation. 

It is well known that statistical methods based on a Gaussian distributional
assumption can be
strongly affected by outliers, and this not different in cluster analysis. 
For fixed $G$, outliers have to be included in a cluster, in turn affecting
their mean and covariance matrix estimators and often the 
classification of many further observations. In order to deal with this,
\cite{BanRaf93} proposed to add a so-called ``noise component'' to the mixture
in order to collect outliers and to prevent them from 
affecting the Gaussian clusters. The density then becomes
\begin{equation}\label{eq:gaussmixnoise}
f(\bx; {\bm{\theta}})=\pi_0\delta+\sum_{g=1}^{G} \pi_g \phi_p(\bx; {\bm{\mu}}_g, {\bf \Sigma}_g),
\end{equation}
$\delta\ge 0,\ \pi_0\in[0,1]$, and now $\sum\nolimits_{i=0}^G \pi_g = 1$.
They proposed to estimate the $\delta$ as $1/M$, where $M$ is the hypervolume of the smallest hyperrectangle to cover all data, assuming that $\delta=0$ outside that hyperrectangle. The number of clusters is still estimated by the BIC, adding the $\pi_0$-parameter to the parameter count. Although this method often works reasonably well, it is actually not the ML estimator for $\delta$ \citep{CorHen11}, and neither is it breakdown robust, because a single extreme outlier can make $M$ arbitrarily large, preventing any other outlier from being classified as noise \citep{Hennig04}. The same holds for another mixture approach that is meant to be more robust than plain Gaussian mixtures, namely mixtures of t-distributions \citep{PeeMcL00}. 

\cite{Hennig04} noted that a method with a better breakdown point can be defined
by fixing $\delta$ in (\ref{eq:gaussmix}). Allowing $\delta$ to be positive
on the whole Euclidean space makes $f$ an improper density,
although a proper density can be defined that constrains the noise component
to occur
in an unspecified set of Lebesgue measure $1/\delta$ that is assumed to 
cover all actually observed data. In this way, 
all other parameters can still be estimated using the EM algorithm, 
enjoy improved robustness 
properties, and observations can still be clustered using (\ref{eq:postprob}). 
For multivariate Gaussian mixtures this has in detail been 
explored by \cite{CorHen16,CorHen17} under the name ``Robust Improper Maximum
Likelihood Estimator'' (RIMLE). \cite{CorHen16} propose to choose $\delta$ as
\begin{equation}
  \label{eq:optdelta}
  \argmin_\delta \left(D(\delta)+\beta\hat\pi_0(\delta)\right),
\end{equation}
where $D(\delta)$ is a measure of the Kolmogorov-type difference between 
the distribution function of 
within-cluster Mahalanobis distances weighted by (\ref{eq:postprob}) between the
observations and the
cluster centre, and the $\chi^2$-distribution function, which should be observed
for perfectly Gaussian distributed observations. The weighting assigns all 
observations to the clusters according to the estimated probability of being
generated by that cluster, which particularly means that observations that have
a high estimated probability of being ``noise'' will be downweighted. Minimising
$D(\delta)$ means that $\delta$ is chosen so that the estimated clusters will 
look optimally Gaussian. This happens if $\beta=0$ is chosen. $\beta$ is a 
tuning constant that allows for tolerating more non-normality within clusters
if in turn the estimated noise probability $\hat\pi_0(\delta)$ is decreased. 
\cite{CorHen16} suggest $\beta=1/3$ as alternative to $\beta=0$.
This is particularly useful for estimating the number of clusters with clusters 
that are not necessarily required to be normal, see Section \ref{sec:simulation}. 

$D(\delta)$ can degenerate and becomes meaningless if $\delta$ is so large 
that all or most observations are classified as noise. Therefore, 
using (\ref{eq:optdelta})
requires that the average posterior
pseudo probability of 
observations to have been generated by the noise component is limited, 
and \cite{CorHen17} propose an
upper bound of 0.5.

Like other methods based on Gaussian mixtures, OTRIMLE needs to address the
issue of a potentially degenerating likelihood due to covariance 
matrices with very small or zero eigenvalues. This is done imposing the constraint
\begin{equation} \label{eq:cov_constraint}
\lmax({\bm{\theta}}) / \lmin({\bm{\theta}})  \leq \gamma < +\infty,
\end{equation}
where $\lmax({\bm{\theta}})$ and $\lmin({\bm{\theta}})$ are the maximum and minimum of the 
eigenvalues of the covariance matrices of the different Gaussian mixture
components parameterised in ${\bm{\theta}}$, and $\gamma\ge 1$ is a constant to be 
chosen by the user. Based on experiments in \cite{CorHen17}, $\gamma=20$ 
seems to be a sensible choice for standardised data (if the measurements of 
different variables
in the data set have different orders of magnitude, there is hardly any 
reasonable way
to specify $\gamma$), although occasionally a user may look for either more 
spherical clusters (which requires smaller $\gamma$) or for even more 
flexibility of the covariance matrices (which requires larger $\gamma$). See
\cite{GEGGIMI18} for a comprehensive discussion of covariance matrix constraints
in Gaussian mixture modelling, {\color{black} and particularly Section 4.1 of \cite{Ritter14}
for robust clustering}. \cite{CGEMI18} argue that the choice of $\gamma$
has impact on the number of clusters, and explore this for the case of a plain
Gaussian mixture model. 

The resulting method is called ``Optimally Tuned RIMLE'' (OTRIMLE), 
and implemented
in the R-package \texttt{otrimle} 
\citep{CRANotrimle}. Theory including consistency
for the canonical functional, a breakdown point, and detailed information
about computation is given in \cite{CorHen17}. A simulation study comparing 
OTRIMLE with plain Gaussian mixtures and alternative robust methods 
is in \cite{CorHen16}. {\color{black} Likelihood-based methods such 
as BIC and ICL should not be used
for estimating the number of mixture components with OTRIMLE, at least not in 
their original form, because the 
parameter $\delta$ is not chosen by ML and affects the comparison of
models fitted for different numbers of components in a way not covered by
likelihood-based theory.}  

\section{An adequacy approach to decide the number of clusters}
\label{sec:approach}
We have argued in the Introduction that the problem of finding a suitable
number of clusters is essentially different from the problem of estimating the
number of mixture components. Even if a Gaussian mixture model is precisely 
fulfilled, a ``submixture'' of several poorly separated Gaussian 
components taken together can still be 
unimodal and even look fairly close to a single Gaussian distribution. In most
applications this would qualify as a single cluster, and the number of 
meaningful real clusters in such a case would be smaller than the number of 
Gaussian mixture components. 

\begin{figure}[htp]
\centering
\includegraphics[width=0.47\textwidth]{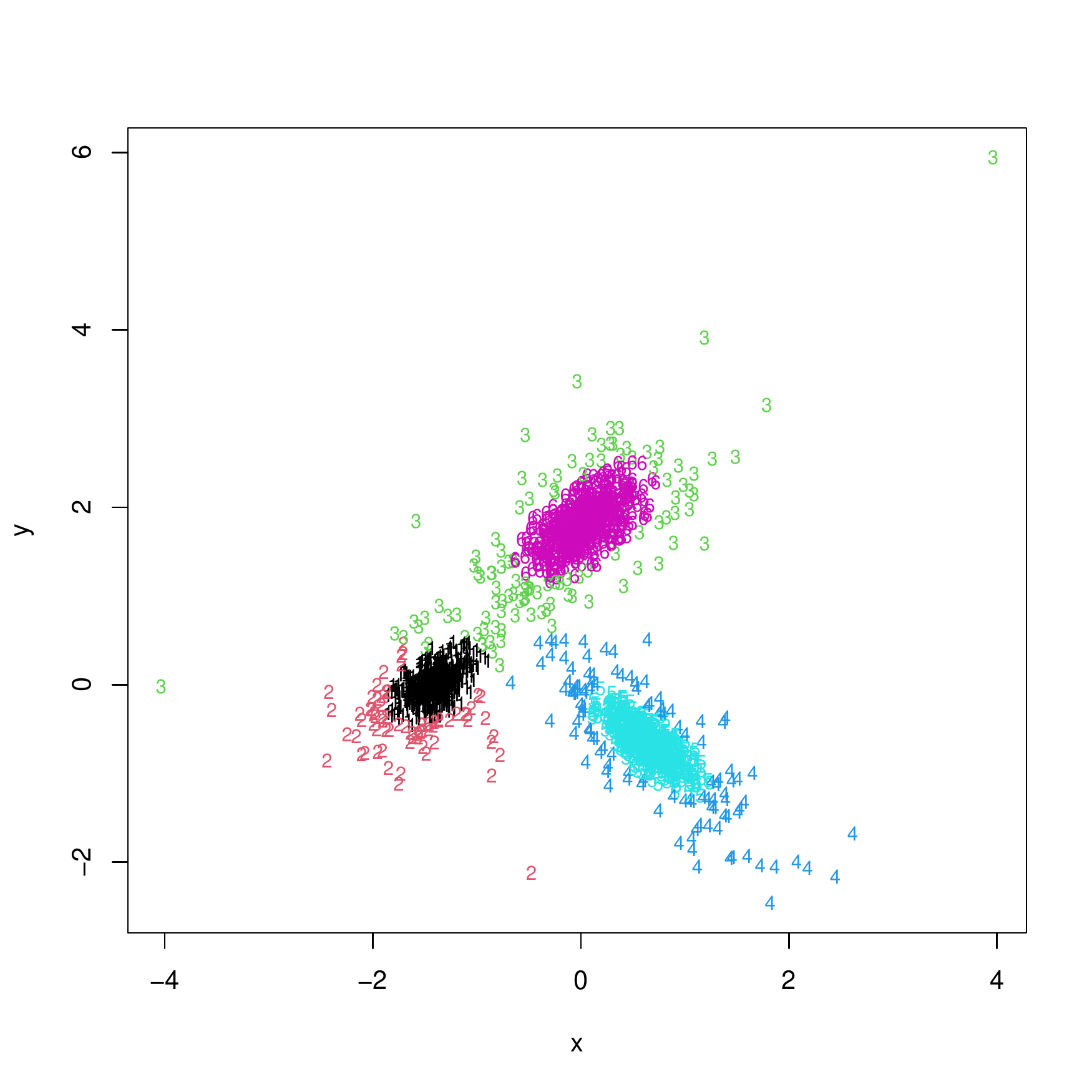}
\includegraphics[width=0.47\textwidth]{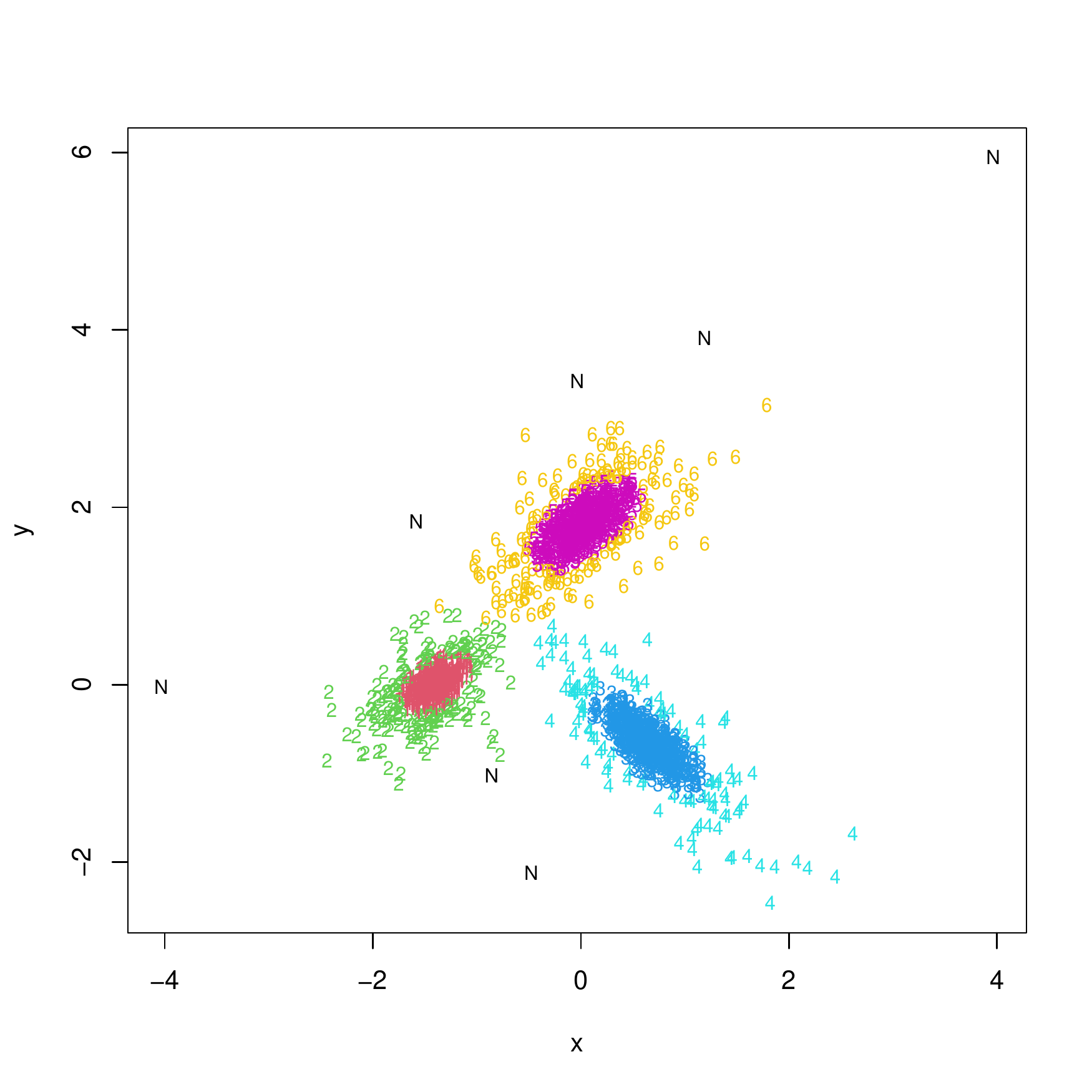}
\caption{\label{fig:t3mclust} Data generated from a mixture of three 
multivariate $t_3$-distributions with clustering by Gaussian mixture fitting
(left side) and Gaussian mixture fitting with noise component (right side);
the number of mixture components was estimated by the BIC.
}
\end{figure}

The problem of estimating the number of Gaussian mixture components is 
ill-posed because any data set generated from a Gaussian mixture with
a certain number of components can be arbitrarily well approximated by a 
mixture with more components. This particularly means that if the 
Gaussian mixture model 
assumption is not precisely fulfilled (as is always the case in reality), with
enough observations a mixture with arbitrarily
many components will fit the data better than a mixture with few components, 
even if the latter may look like an excellent representation of the intuitive 
clusters in the data. This is illustrated in Figure \ref{fig:t3mclust}, which
shows data generated by a mixture of three multivariate $t_3$-distributions
(generated by the setup ``TGauss.3l'' in \cite{CorHen16}). The left side shows
a clustering from a plain Gaussian mixture produced by \texttt{mclust} with
default settings. Although there are three elliptical clusters clearly visible,
the BIC estimates the number of Gaussian mixture components as 6, because the
intuitive clusters have not been generated exactly by a Gaussian distribution.
Adding a uniform noise component (right side of Figure \ref{fig:t3mclust})
classifies some outliers appropriately as ``noise'', but does not help with the
estimation of the number of clusters, as the BIC still estimates 6 Gaussian
components. A mixture of t-distributions will fit these data well with three
mixture components, however if the underlying distributions are not exactly 
t-distributions, it runs into similar problems, see Section 
\ref{sec:simulation}. A consistent method
such as the BIC has more use for picking a mixture that fits the empirical
density well than for interpreting the resulting components as clusters.   
  
This implies that the problem of deciding the number of clusters is not a well
defined statistical estimation problem. It does not only rely on parameters of
an assumed underlying distribution, but also on user decisions. Even assuming 
that the Gaussian distribution is used as a ``cluster prototype'', i.e., a 
cluster should look Gaussian or similar,
the user has to decide
\begin{enumerate}
\item what is required of a data subset to be interpreted as cluster,
\item how far from a Gaussian distribution a within-cluster distribution is
tolerated to be,
\item in case that some observations can be classified as outliers/noise, how
small and homogeneous an outlying data subset is required to be in order
to be interpreted as cluster rather than a group of outliers. 
\end{enumerate}
These decisions cannot be made from the data alone, and therefore user tuning
is essential for estimating the number of clusters. We believe that this is 
quite generally the case in cluster analysis, and that the vast majority of the
literature ignores this, probably because most users expect a solution without
having to make decisions, and a solution that depends crucially on user tuning
may not be accepted as ``objective''; see \cite{GelHen17} for a discussion of
this issue.

We now introduce a general scheme for deciding the number of clusters that can
be applied to general model-based clustering methods, and that can be tuned by
the user addressing the issues above.

The scheme is based on a general approach to model selection proposed first
in \cite{Davies95} and more explicitly (in the context of nonparametric 
regression) in \cite{DavKov01}. The idea is that one can choose the simplest 
model that is adequate for the data in the sense that it produces data that
cannot be distinguished from typical data generated by the model. Obviously,
more complex models can be adequate as well, as is the case in mixture 
modelling, but a more complex model will not be chosen if a simpler one exists
that is already adequate. Entry points
for user tuning are: 
\begin{enumerate}
\item the target model, i.e., the model for which adequacy of the data
is evaluated (in cluster analysis this will often be a mixture model; here
a Gaussian mixture model, as we assume that the Gaussian distribution serves
as ``cluster prototype''),
\item the statistic, or potentially more than one statistics, that are used 
to distinguish the data from what is expected under the model 
(in cluster analysis a statistic $Q$ is required that measures whether what is
interpreted as clusters behave as clusters should behave in the application 
at hand), 
\item how atypical data has to look like in order to decide against the model
(standard significance levels such as 0.01 or 0.05 may be used),
\item the formal definition of simplicity $S$ (in cluster analysis the standard 
choice would be the number of clusters, but we will penalise this with the
estimated noise proportion in order to stop the method from declaring too many
observations ``noise'').
\end{enumerate}
We will work with a statistic $Q$ that does not allow for simple 
analytic derivation of its distribution for data generated by a mixture, and
therefore its distribution will be approximated by parametric bootstrap.

Let $\bX=(\bx_1,\ldots, \bx_n),\ \bx_i=(x_{i1},\ldots,x_{ip})^\top
\in\R^p,\ i=1,\ldots,n$ be the data set
and $C_G(\bX)$ be the output of the clustering method $C$ with $G$ clusters 
on $\bX$. 
Here is the general scheme:
\begin{enumerate}
\item Choose a target model, a clustering method that fits the target model,
a statistic $Q$ that 
measures clustering quality, and a statistic $S$ measuring the simplicity 
of a fit. In practice also a maximum number $G_{max}$ of clusters and a 
number of bootstrap resamples $B$ are required.
\item For $G=1,\ldots,G_{max}$, compute a fit (clustering) 
of $\bX$ with $G$ clusters.
\item For $G=1,\ldots,G_{max}$, generate $B$ data sets $D_{G,b},\ b=1,\ldots,B$ 
from the fitted model.
\item For given $G$, the clustering is adequate for the data if $Q(C_G(\bX))$ is
consistent with the empirical distribution of $Q(C_G(D_{G,b}))$, see Section 
\ref{sec:bootstrap}. 
\item The final number of clusters is chosen as 
$\argmin_{G \mbox{ adequate}} S(G)$. In the 
simplest case $S(G)=G$, and the scheme can be stopped once an adequate $G$
is found.   
\end{enumerate}
A possible outcome of the scheme is that no clustering is adequate.
This is informative for the user in its own right,
and means that the data are not compatible 
with the target model, at least not for $G\le G_{max}$. 
There are various options to enforce a clustering if it is required anyway.
One could try a larger $G_{max}$, choose the best found clustering according to 
$C(G)$, or $\frac{C(\bX)-m_{QG}}{s_{QG}}$ (see Section \ref{sec:bootstrap} 
for the definition), or try a non-model based clustering method.

\section{Key decisions and tuning} \label{sec:statistic}
{\color{black} We illustrate the general approach by using it for deciding the number of clusters with OTRIMLE.} The number of bootstrap
replications $B$ and the maximum number of clusters $G_{max}$ should
optimally be as large as possible, but the method is computationally 
intensive, so they need to be limited for pragmatic reasons. The
choice of $G_{max}$ should also depend on potential background information
about a realistic or required number of clusters.  $B$ should be at least 
around 20 to  give the method some stability, but $B=100$ and higher would 
be better. The further choices are less straightforward.
\subsection{Data generation from the target model}
The target model {\color{black} in case of OTRIMLE}
should be a Gaussian mixture with noise, similar to 
(\ref{eq:gaussmixnoise}), but (\ref{eq:gaussmixnoise}) in the given form
is not a proper probability model without constraining the set 
where noise (i.e., observations from mixture 
component zero) can occur.   

With all parameters estimated by OTRIMLE and assuming the noise to be
constrained to an unspecified set of Lebesgue measure $1/\delta$, the
estimated posterior probability of observation $\bx_i,\ i=1,\ldots,n,$ 
to be noise is
\begin{displaymath}
  \hat p_{i0}=\frac{\hat\pi_0\delta}
{f(\bx_i; \hat{\bm{\theta}})}.
\end{displaymath}
For data generation from the target model for the parametric bootstrap, an 
observation is assigned to the noise with probability $\hat\pi_0$, and given 
that it is assigned to the noise, we propose to resample it from the 
existing data set with the noise distribution defined by
\begin{displaymath}
  \hat P_0\{\bx_i\}=\frac{\hat p_{i0}}{\sum_{h=1}^n \hat p_{h0}},
\end{displaymath}
so that the probability of every observation to be drawn as noise is 
proportional to its estimated probability to be noise in the data set.
Non-noise data are generated in a standard way from the estimated Gaussian
mixture.
\subsection{The clustering quality statistic}
The clustering quality statistic $Q$ is meant to formalise what a ``good'' 
clustering is. We do not insist on a precisely Gaussian shape, but we assume 
that the clusters of interest here should be elliptical and unimodal 
with density decreasing from the mean symmetrically in all directions. In such
a case the use of the Gaussian distribution as a cluster prototype and the 
Gaussian mixture approach seem justified. 

The $Q$ proposed here measures in a nonparametric way to what extent the 
clusters have such a shape. We start from a one-dimensional measure for a single
cluster. {\color{black} For $p>1$, within-cluster principal components (PCs)
are computed, and 
the values of the one-dimensional measure} 
are then aggregated over all PCs and over all clusters to compute the overall $Q$. The definition is
not motivated by any model-based optimality theory, but rather custom-made
in order to express exactly what is required. It is based on a test for 
unimodality by \cite[p. 79]{Pons13}. 

\begin{figure}[htp]
\centering
\includegraphics[width=0.33\textwidth]{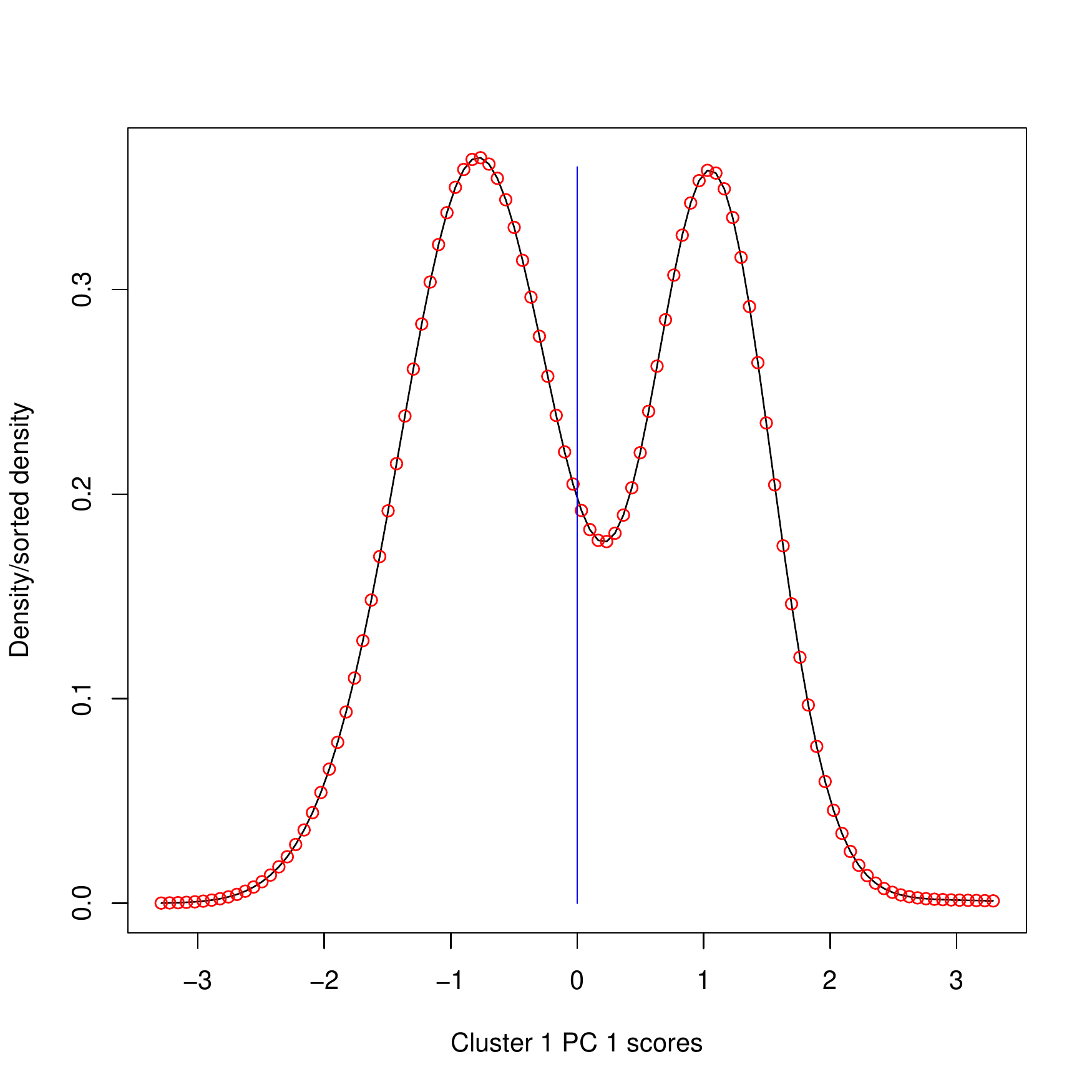}
\includegraphics[width=0.33\textwidth]{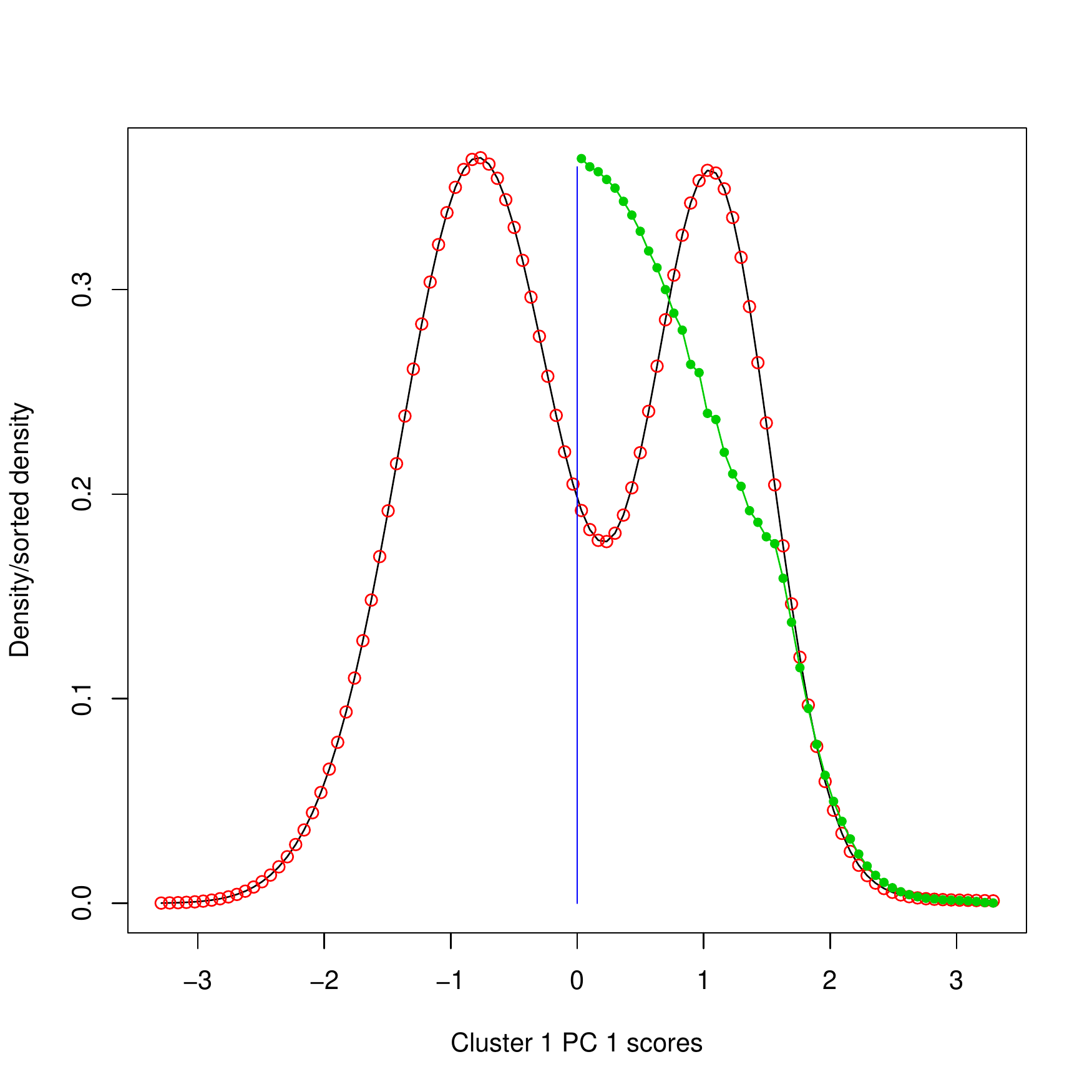}
\includegraphics[width=0.33\textwidth]{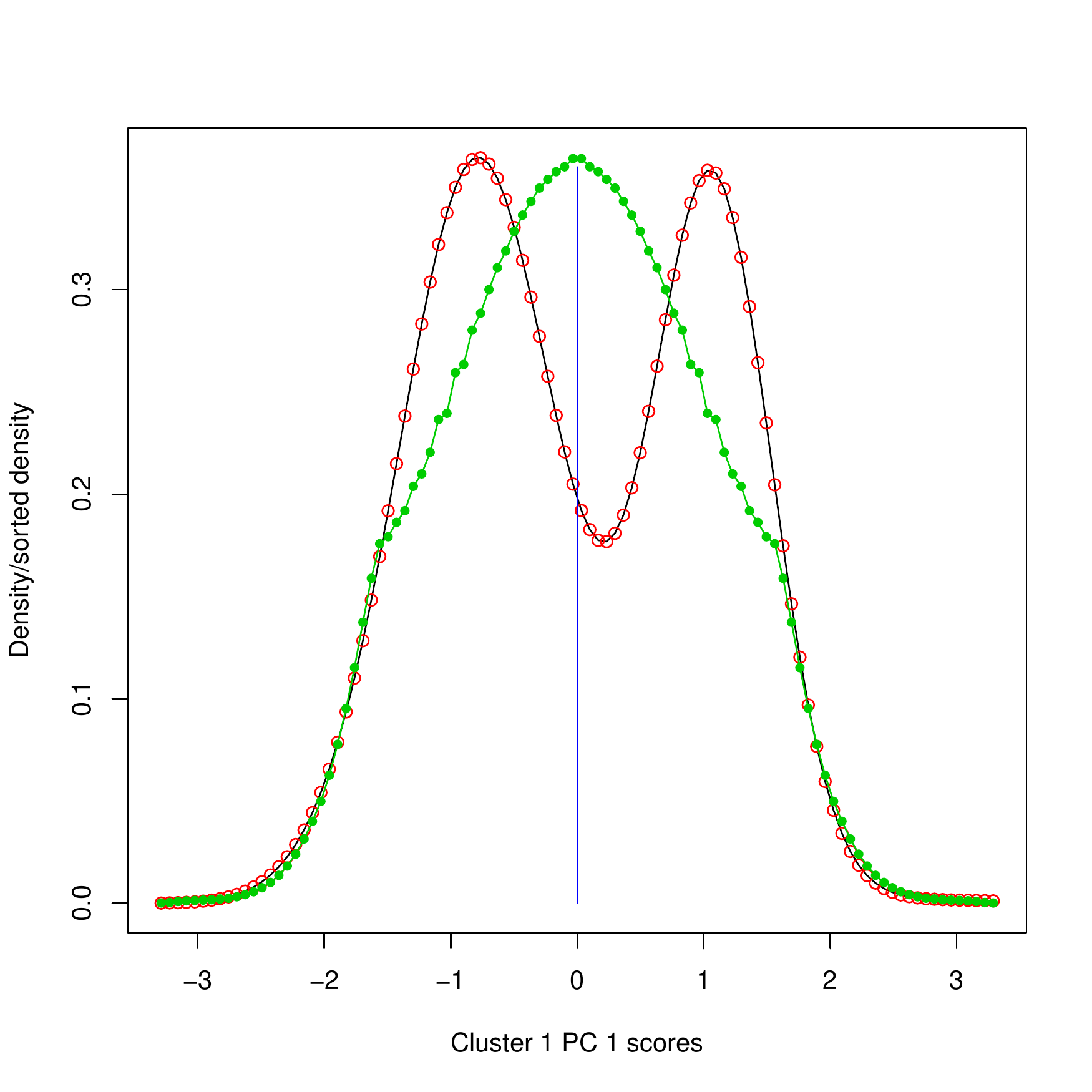}
\caption{\label{fig:densities} Illustration of the one-dimensional measurement
of cluster quality. Left side: Suppose this is the kernel-estimated density 
for the weighted data set within the first estimated cluster at 
$z_1,\ldots,z_q$, obviously not looking unimodal. 
Middle: Density values at $z_1,\ldots,z_q$
are ordered from the largest to the smallest. Pairs of density values 
(the two largest ones, then the third and fourth largest and so on) are 
averaged, and the resulting density values are shown on the right side of the
mean at $z_{q/2+1},\ldots,z_q$ from largest to smallest. Right side: The same
values are also put on the left side of the mean in descending order from the 
mean to the outskirts, producing a density symmetric about the mean. $\tilde Q_{g}$ is
the root of the averaged squared difference between these.}
\end{figure}

Assuming one-dimensional data standardised to have mean zero and variance one
in cluster $g=1,\ldots,G,$ we use the following definition:
\begin{enumerate}
\item Choose a kernel density estimator and $q$ 
points $z_1<z_2<\ldots<z_q$
symmetrically around the mean. Our software uses the default of the 
R-function \texttt{density}, $q=100$,
and the 100 points are chosen as $p$-quantiles of the standard Gaussian
distribution with $p$ ranging from 0.005 to 0.995 in equidistant manner.  
\item Compute kernel density estimators at the quantiles 
$\hat f(z_1),\ldots, \hat f(z_q)$ based on a weighted sample in which 
$x_{ij}$ has a weight according to (\ref{eq:postprob}).
\item Let $\hat f^{(1)}\ge \hat f^{(2)}\ge \ldots\ge \hat f^{(q)}$ be the sorted
version of $\hat f(z_1),\ldots, \hat f(z_q)$. 
\item For $h=1,\ldots,q/2$, let $\hat f^{*h}=\frac{\hat f^{2h-1}+\hat f^{2h}}{2}$.
This implies that $f^{*1}, f^{*2},\ldots,f^{*(q/2)},f^{*(q/2)},\ldots,f^{*1}$ is
a symmetric version of the original  $\hat f(z_1),\ldots, \hat f(z_q)$.
\item Compare the symmetrised kernel density with the mean ($q_l$ and $q_r$
refer to the left and right side of the mean, respectively):
  \begin{displaymath}
    q_l=\sum_{i=1}^{q/2}(\hat f(z_{q/2+1-i})-\hat f^{*i})^2,\ 
q_r=\sum_{i=1}^{q/2}(\hat f(z_{q/2+i})-\hat f^{*i})^2.
  \end{displaymath}
Aggregating: $\tilde Q_{g}=\sqrt{\frac{1}{q}(q_l+q_r)}$. 
\end{enumerate}
The process is illustrated in Figure \ref{fig:densities}.
In case that the estimated density in fact decreases monotonically and 
symmetrically from the mean, $\tilde Q_{g}=0$, which is the best possible value.

For aggregating $\tilde Q_{g}$-values over different clusters, it is 
important to
take the size of the estimated clusters, i.e., $\hat\pi_g,\ g=1,\ldots,G,$ into
account in order to avoid that the overall measure is dominated by a highly
unreliable value from a small clusters. The rationale is not to give bigger 
clusters more weight, because this is about estimating the number
of clusters, so small clusters that are bad should not be tolerated. However,
$\tilde Q_{g}$ can also be expected to be more variable for even valid small 
clusters, and this needs to be accounted for. Therefore we use
\begin{displaymath}
Q_{g}^*=\frac{\tilde Q_{g}-E_{n\hat\pi_g}\tilde Q_{g}}{\sqrt{\var_{n\hat\pi_g}(\tilde Q_{g})}},
\end{displaymath}
where the expectation $E_{m}$ and variance $\var_m$ are computed assuming $m$
i.i.d. observations from an ${\cal N}(0,1)$-distribution in the corresponding
cluster. These values can be simulated to very high precision and interpolated
to allow for non-integer $m$. {\color{black} This idea is similar to ``pivoting'' in bootstrap
inference \citep{Hall92}.}

For $p$-dimensional clusters with $p>1$, within-cluster PCs
are computed first, based on the weighted within cluster data with weights 
according to (\ref{eq:postprob}) again. For $j=1,\ldots,p$, let $Q_{jg}$ 
be $Q_g^*$ computed on the $j$th standardised within-cluster PC 
of cluster $g$. Aggregating information
from the PCs,
\begin{displaymath}
  Q_g=\frac{1}{p}\sum_{j=1}^p(Q_{jg}^2){\bf 1}(Q_{jg}>0),
\end{displaymath}
where ${\bf 1}(\dot)$ denotes the indicator function. The rationale here is that (a)
if $Q_{jg}\le 0$ it means that on the
$j$th PC, the symmetric unimodality statistic behaves as expected under a
Gaussian distribution or even better, so there is no indication whatsoever 
against this being a cluster, and (b) squaring positive $Q_{jg}$ will 
emphasise problematic issues in certain PCs. {\color{black} The contribution 
of the first PCs is not upweighted in the definition of $Q_g$, because 
potential 
issues with unimodality are of interest along all PCs in the same way, although
one could intuitively expect that issues occur more often along the
first PC.}

Finally, for the same reason squares are applied when aggregating over 
the clusters
in order to make $q$ sensitive against substantial issues in any cluster:
\begin{displaymath}
  Q(G)=\sqrt{\sum_{g=1}^G(Q_g^2)}.
\end{displaymath}
{\color{black} A number of alternatives choices of $Q$ could be considered. 
Other tests for
unimodality have been proposed \citep{Silverman81,HarHar85,SiFoTeLa18}. $Q$ as
defined above was chosen because it does not only measure unimodality but also
symmetry (elliptic shape is measured to some extent by assessing the symmetric 
shape in all PC directions), 
and it allows for relatively straightforward aggregation over 
different mixture components and dimensions, because it measures
deviations from the symmetric unimodal shape directly. Measuring unimodality
for multivariate data is hard, and it cannot be ruled out
that violations are only apparent in directions other than the PCs. 
Multimodality can often
be expected to lead to increased variance \citep{SiFoTeLa18}, and therefore the
first PCs are good candidates for detecting it, but exceptions exist. 

It may be possible to adapt other unimodality 
statistics to our approach as well. Also, $Q$ could be defined in different ways
to measure different cluster characteristics of interest. As a simple variant 
the symmetry requirement can be dropped by centering step 5 at the 
maximum estimated
density mode rather than the mean, and compare the estimated data density with
the un-symmetrised sorted density values on the left and right side of the 
mode, respectively, which is the original proposal by \cite[p. 79]{Pons13}.
$Q$ could also take into account classification entropy as the ICL does.
More than one statistic can be employed at the same time to measure multiple 
features of the clusters \citep{AkhHen20}. Elaboration of these ideas 
is left to future work.}

\subsection{Bootstrap adequacy}\label{sec:bootstrap}
Because the method
is computer intensive and precise quantiles may require a too large $B$, 
$G$ will be defined to be adequate if 
\begin{equation}\label{eq:qualitycut}
\frac{Q(G)-m_{QG}}{s_{QG}}\le c,
\end{equation} 
where $m_{QG}$ and $s_{QG}$ are 
location and scatter 
statistics of the empirical distribution of $Q(G)$ for data generated
from the fitted model. We have observed that with OTRIMLE (as potentially 
with other clustering methods) 
$Q(G)$ may produce outlying values. Certain fitted distributions
may generate data sets that are quite ambiguous regarding the optimal 
clustering and the number of clusters. Such outlying values normally indicate 
a very bad clustering, and $Q(G)$ on the original data set
should not be assessed as adequate just
because certain $Q(G)$ on bootstrapped data are even worse. 
For this reason, $m_{QG}$ and $s_{QG}$
should be chosen robustly. We suggest the robust $\tau$-estimator for
location and scale \citep{MarZam02}. With appropriate consistency factor,
this is consistent if the parametric bootstrap distribution of $Q(G)$ is 
Gaussian {\color{black} (we currently do not have a proof for this, so this
is just heuristic; Chebyshev's inequality can be used for guidance with general
distributions)}, 
allowing for a standard interpretation of the constant $c$. 

$Q$ is assumed to be defined
so that lower values imply a better clustering quality, and adequacy will only
be rejected if $Q(G)$ is too large. 
Choosing, e.g., $c=2$ then means that if $Q(G)$ on bootstrapped data follows a 
Gaussian distribution,
the probability that adequacy is rejected is about 0.023.
\subsection{The simplicity measure}\label{sec:simplicity}
The simplest choice for the simplicity measure $S$ is $S(G)=G$; a model is
seen as simpler if it has fewer clusters. This is appropriate for standard
non-robust clustering, but it is problematic if it is allowed to classify
a number of observations as ``noise''. With OTRIMLE, as well as with trimmed
clustering and the noise component in \texttt{mclust}, it would be possible 
to declare all observations ``noise'' that make clustering ambiguous or belong
to small clusters, in which case a high quality clustering with small $G$ 
for the remaining observations could be found easily. For this reason, and
because it is generally ambiguous whether observations that belong to small 
groups in some distance from the bigger clusters should be declared noise or
clusters on their own, too much noise should be penalised. We propose
\begin{equation}\label{eq:optg}
  S(G)=G+\frac{\hat\pi_0}{p_0},
\end{equation}
where $p_0$ is a constant chosen by the user. It specifies the smallest 
percentage of additional noise that the user is willing to trade in for 
adding another cluster, i.e., if $p_0=0.05$ (which we use as a default),
it means, say, that a clustering with $G=6$ and $\hat\pi_0=0.04$ is assessed
as ``simpler'' as a clustering with $G=5$ and $\hat\pi_0=0.1$. The former
clustering will then be preferred by our method if both clusterings are 
adequate. Particularly this will normally imply that clusters with 
$\hat\pi_0<p_0$ are not found, because they could simply be declared noise and the
resulting clustering would be ``simpler'' and as adequate, although there may be
exceptions in case that the smallest cluster has a high quality $Q_g$
compared to the other clusters.
\section{Experiments}\label{sec:experiments}
The adequacy approach to choose the number of clusters with OTRIMLE (called 
AOTRI in the following) is compared to different mixture model-based
methods in a simulation study and on two data sets of scientific interest, 
one with and the other one without given true $G$. There is always a tension 
between stating that a method requires user tuning dependent on the specific 
situation, and running it in a default fashion on artificial data sets, but we 
think that both of these have their justification. Where user decisions can be
used with convincing justification to adapt the method to what is required in a 
given application, this is certainly recommended. However, in many situations 
the user does not have a clear idea how to make some or all of these choices, 
and therefore defaults are often useful. They are also required in order to
compare the method in a ``neutral'' fashion with others. In the following we 
choose $p_0=0.05$ in (\ref{eq:optg}), 
i.e., we prefer a solution with one cluster more if that
reduces the estimated noise by 0.05 or more. We did some experiments
with $p_0=0.02$ (not shown), but results were rarely different. 
We choose 
$c=2$ in (\ref{eq:qualitycut})
as maximum value of the standardised clustering quality for the model 
to still count as ``adequate''. The maximum eigenvalue ratio for covariance 
matrices was chosen as $\gamma=20$. Variables 
were standardised before clustering in order to allow for a 
scale-independent interpretation of 
$\gamma$ except where mentioned explicitly.

We looked at both $\beta=0$ and $\beta=1/3$ in
(\ref{eq:optdelta}), the latter called AOTRIB, and 
meaning that for fixed $G$ more non-Gaussianity
within clusters is tolerated if that reduced noise. Results were occasionally
different. Note that $\beta$ trades non-Gaussianity against 
noise for fixed $G$, whereas $c$ tunes trading non-Gaussianity against non-adequacy of the
non-noise,
usually leading to a larger $G$ (if anything changes at all, which it often 
does not). 

We chose $G_{max}=10$ in the simulations. This choice 
does not  matter, however, as long as the finally chosen $G$ has a value of
$S(G)<G_{max}+1$ in (\ref{eq:optg}), because then it will be chosen regardless
of results for higher $G$. As far we have seen, for all data sets, larger 
$G_{max}$ could not have changed results for this reason; for the BIC 
{\color{black} and ICL}
this can never be known, which is an advantage of our approach.

The number of bootstrap replicates is chosen as
$B=30$ in the computer intensive simulations, but $B=100$ in Sections 
\ref{sec:srna} and \ref{sec:simdata}.

{\bf Declaration of selection bias.} As this paper introduces a new method,
as a proof of concept we need to show some situations in which it works well.
We looked at some other data sets and data generating mechanisms 
(although usually with a very small number of test runs). In many cases there
was no big difference between the different methods, and sometimes 
\texttt{mclust} with or without noise, or a mixture of t-distributions or 
skew t-distributions worked better, though never all of them. Sometimes nothing
worked well. So we do not claim that AOTRI/AOTRIB is universally the best, 
just where we show it is. {\color{black} Data generating process (DGP)} 3 was the
first DGP we tried, and we show it despite 
not being a clear win for the new methods.  
\subsection{Simulation study}\label{sec:simulation}
In this study we compare AOTRI and AOTRIB with some 
mixture model-based clustering methods that estimate the number of clusters 
using the BIC {\color{black} or the ICL}. 
More precisely, we use the R-package \texttt{mclust} with default settings
for fitting
a Gaussian mixture {\color{black} with BIC and ICL 
(GBIC, GICL), and with noise component (GNBIC);} the noise component is
initialised by the R-function \texttt{NNClean} in package \texttt{prabclus} 
with parameter \texttt{nnk=5} \citep{ByeRaf98}. We use the R-package 
\texttt{teigen} \citep{AnWiBoMc18} 
for fitting mixtures of multivariate t-distributions using the
BIC and ICL (TBIC, TICL). We use the R-package
\texttt{EMMIXskew} for fitting mixtures of skew $t$-distributions
(\cite{WaNgMc09,LeeMcL13}; SKTBIC). 
We use fully flexible covariance matrices 
and degrees of freedom if possible, but 
sometimes \texttt{EMMIXskew} does not deliver a solution with the default 
settings, in which case we try out more constrained covariance matrix models as
offered by \texttt{EMMIXskew} until a valid solution is found, which in the 
simulations ultimately always was the case. 100 data sets have been generated 
from each DGP.

\begin{figure}[htp]
\centering
\includegraphics[width=0.45\textwidth]{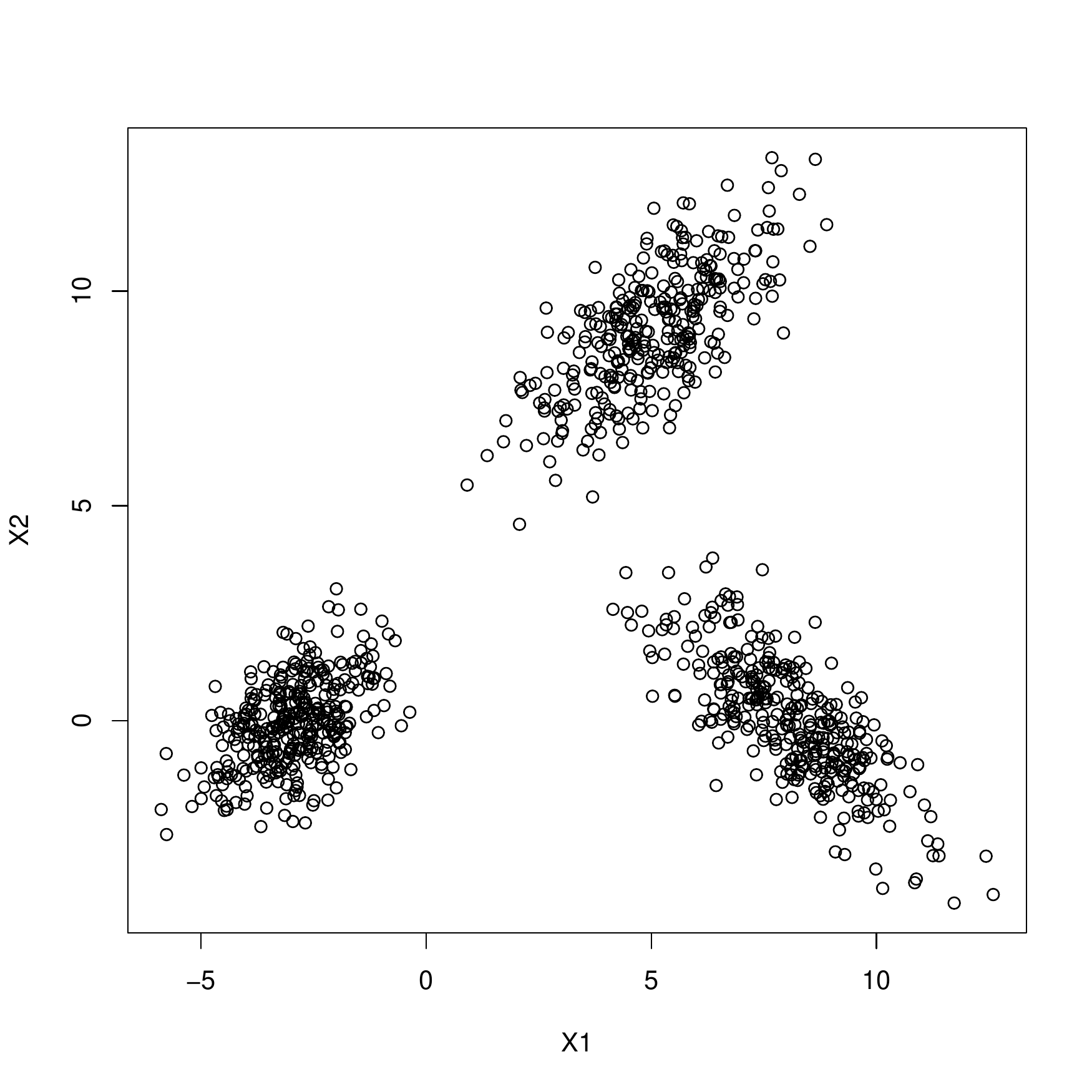}
\caption{\label{fig:noiseless3h} First two dimensions of data from simulated
DGP 1, generated from a mixture of three 
multivariate Gaussian distributions.
}
\end{figure}

{\color{black} 
We 
consider the chosen number of clusters and the adjusted Rand index comparing
the resulting clustering with the true clustering 
(ARI; \cite{HubAra85}). 
This becomes 
1 for perfect correspondence, and 0 is its expected value for comparing
two random clusterings. For the AOTRI variants and GNBIC the noise
component is included as a cluster in the computation of the ARI with one 
exception. In a real situation classifying observations as ``noise'' indicates
that cluster membership is unclear, and on this basis these observations 
could be excluded from the computation of the ARI, but this could be seen as an
unfair advantage for these methods, because the other methods are assessed 
based on all observations including those that are hardest to classify. 
Therefore we decided to include the noise in the ARI computation, although
we show both results for DGP 4. 

We simulated data from four DGPs. DGP 1 is a plain Gaussian mixture 
with $G=3$ components, all with probability
$\frac{1}{3}$. There are $n=1000$ observations in $p=10$ dimensions. The means
and covariance matrices of the first two dimensions (see Figure 
\ref{fig:noiseless3h}) are:
$\mu_1=(-3,0)'$, $\Sigma_1[1,1]=\Sigma_1[2,2]=1$, $\Sigma_1[1,2] =  0.5$; %
$\mu_2=(8,0)'$, $\Sigma_2[1,1]=\Sigma_2[2,2]=2$, $\Sigma_2[1,2] = -1.5$  %
$\mu_3=(5,9)'$, $\Sigma_3[1,1]=\Sigma_3[2,2]=2$, $\Sigma_3[1,2] =  1.3$. 
Each of variables 3-10 is generated from ${\cal N}(0,1)$ independently of 
the others. Standardising the variables implies downweighting of the cluster 
structure compared to the non-informative variables 3-10, causing problems for
all methods including GBIC and GICL of which the model assumptions are fulfilled
here, and therefore for DGP 1 and DGP 2 variables were not 
standardised before clustering.

\begin{figure}[htp]
\centering
\includegraphics[width=0.48\textwidth]{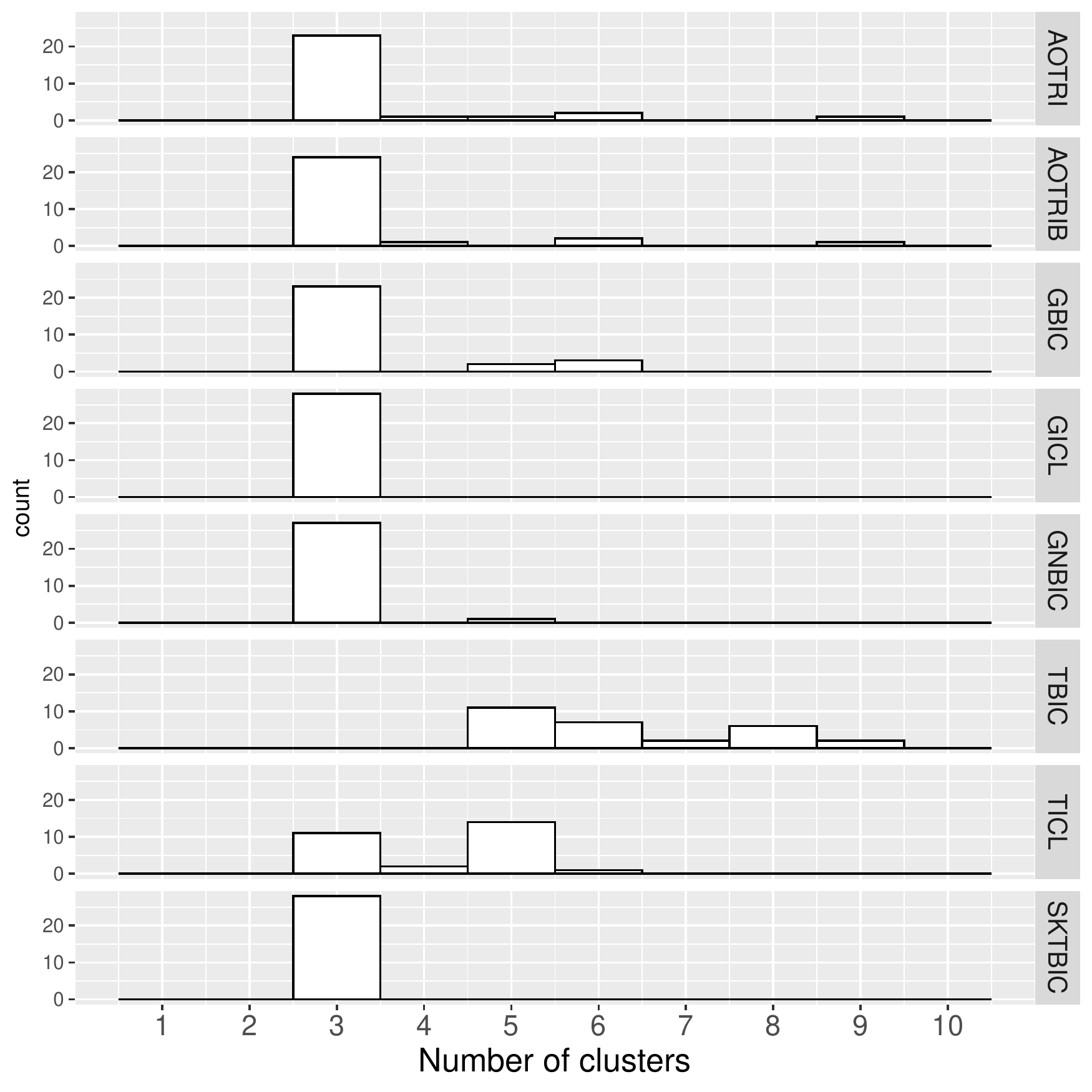}
\includegraphics[width=0.48\textwidth]{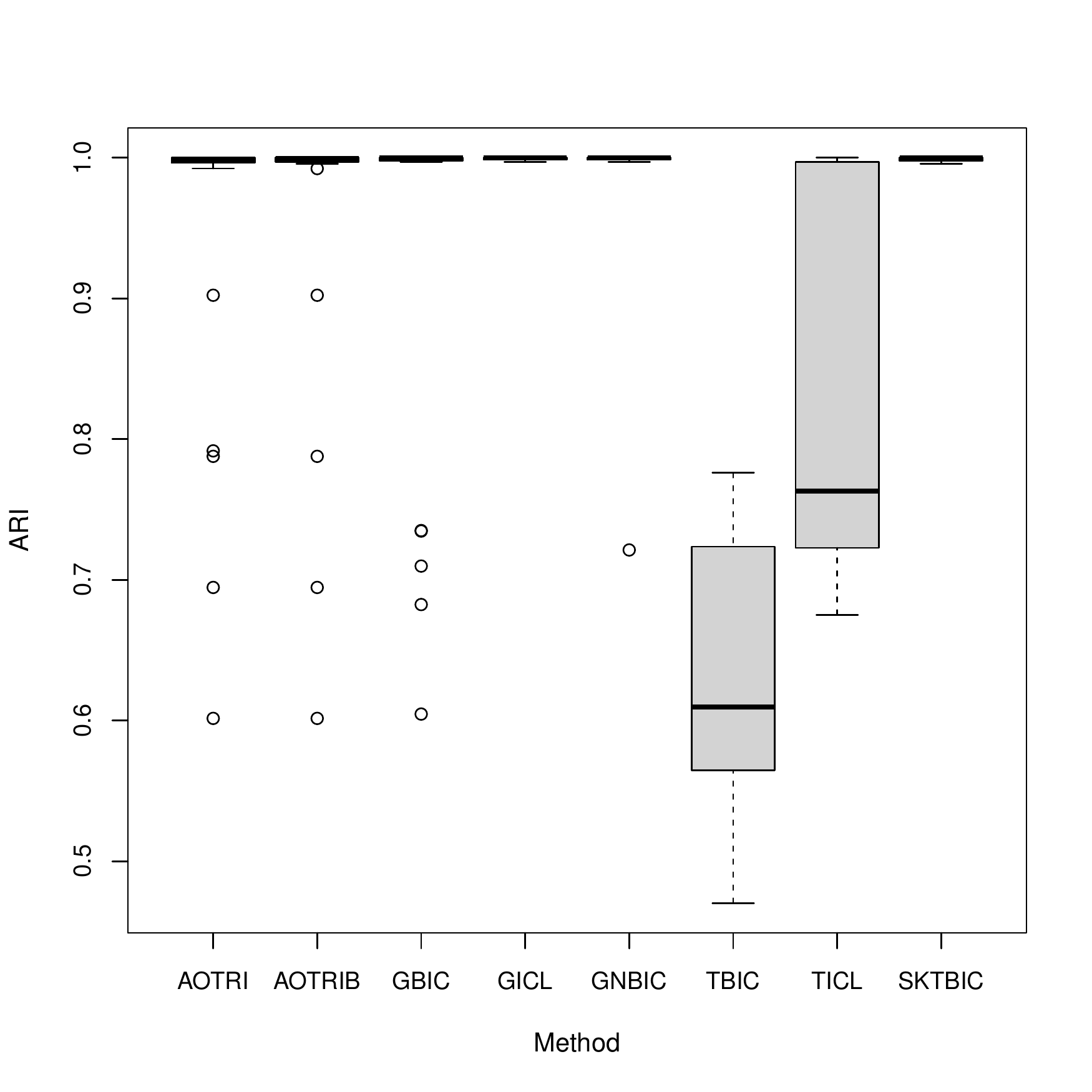}
\caption{\label{fig:noiseless3hresults} Left side: Distribution of numbers of
clusters by method for DGP 1 (true $G=3$) over 100 simulation runs. Right side:
Corresponding distributions of adjusted Rand index values. 
}
\end{figure}

The results for DGP 1 are shown in Figure \ref{fig:noiseless3hresults}, see
Table \ref{tab:arimean} for the mean ARI values. GICL produces perfect results
here, as does, somewhat surprisingly, SKTBIC. TBIC and TICL perform 
substantially worse, TICL being the better of the two. What happens here is 
analogous to the problem shown in Figure \ref{fig:t3mclust}; as the clusters
are not t-distributed, the methods often add mixture components to approximate
the Gaussian distributions better by t-distributions. 
The remaining 
methods including the two AOTRI methods 
almost always find the correct clustering, with a few exceptions. 

\begin{figure}[htp]
\centering
\includegraphics[width=0.48\textwidth]{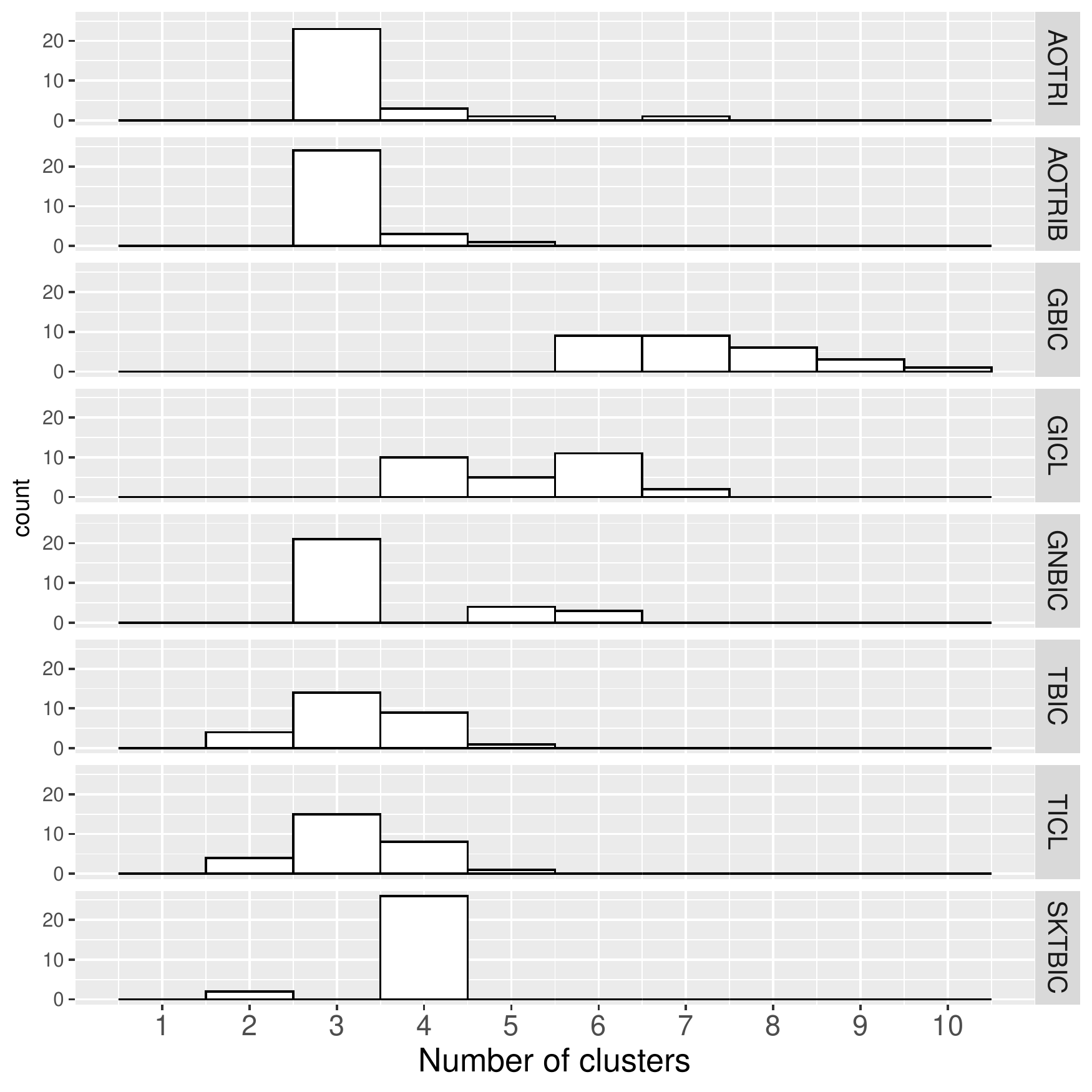}
\includegraphics[width=0.48\textwidth]{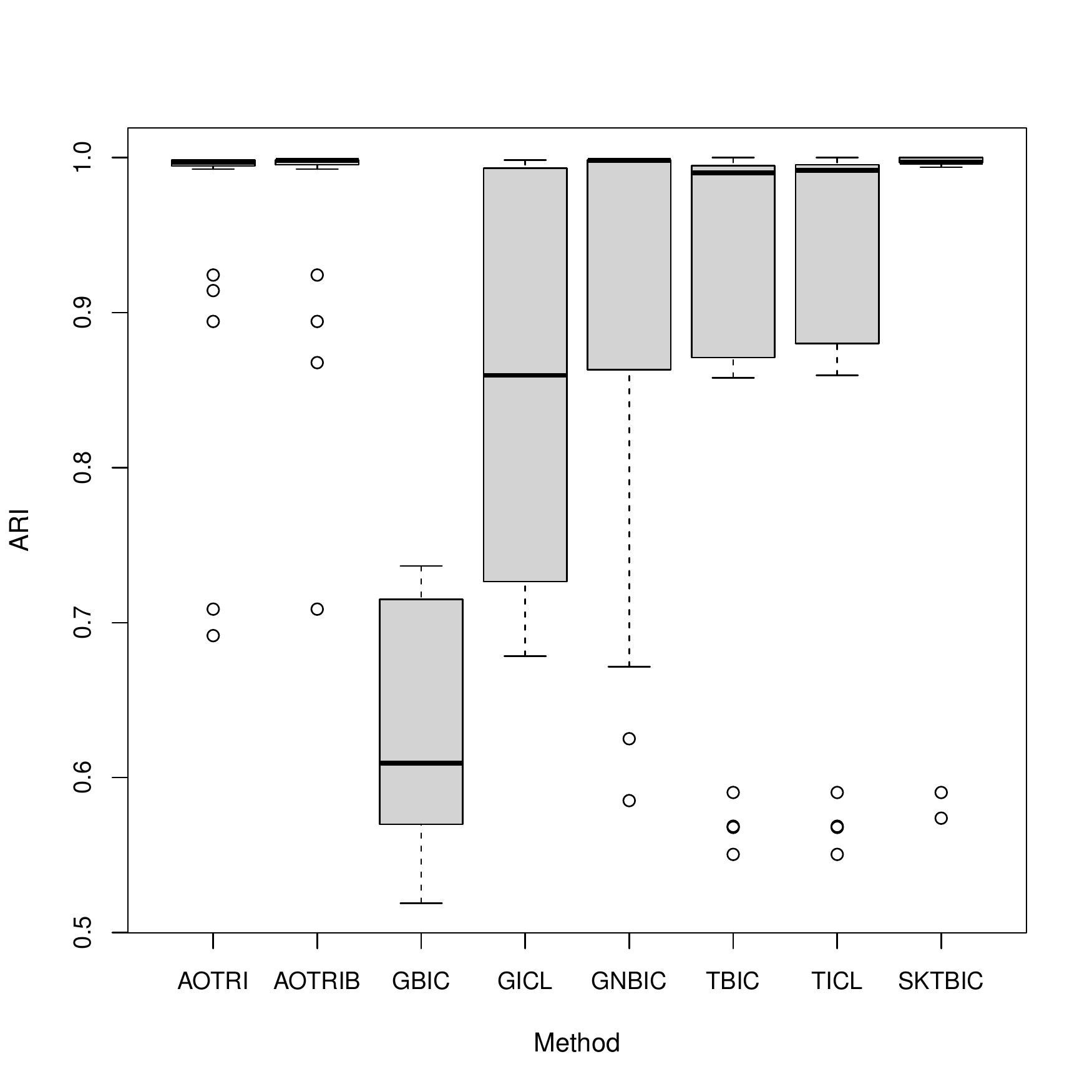}
\caption{\label{fig:noiselessoutresults} Left side: Distribution of numbers of
clusters by method for DGP 2 (true $G=3$) over 100 simulation runs. Right side:
Corresponding distributions of adjusted Rand index values. 
}
\end{figure}

DGP 1 serves as a baseline for DGP 2, which is identical to DGP 1, except that
one observation from 
cluster 1 has its value in the third variable replaced by 1000, and therefore is now a gross outlier. The results for DGP 2 are shown in Figure 
\ref{fig:noiselessoutresults} and Table \ref{tab:arimean}. The AOTRI variants
produce even slightly better results than in DGP 1; the added extreme outlier 
may occasionally stop truly Gaussian observations from being classified as 
noise. The outlier also seems to regularise TBIC and TICL to some extent, 
although their results are still not excellent. 
The result of SKTBIC is still very good, typically adding a cluster for the
outlier alone, and the robustness of GNBIC 
looks satisfactory. On the other hand, GBIC deteriorates strongly, and GICL 
worsens significantly as well. The effect of the outlier on these methods is
in many cases that the choice of the covariance matrix model is affected and
a model is used that requires more mixture components to fit the clusters.}

\begin{table}[htp]
\caption{\label{tab:arimean}
Average adjusted Rand index values over 100 simulation runs. The last line
gives the values for DGP 4 excluding the observations that were classified as
noise.}
\centering
\begin{tabular}{|l|r|r|r|r|r|r|r|r|} \hline
Method & AOTRI & AOTRIB & GBIC & GICL & GNBIC & TBIC & TICL & SKTBIC\\ 
ARI         &        &       &       &       &   &&&    \\
   \hline
DGP 1         & 0.955 & 0.963 & 0.945 & 0.999 & 0.989 & 0.631 & 0.842 & 0.999\\
DGP 2         & 0.967  & 0.976 & 0.632 & 0.844 & 0.918 & 0.901 & 0.906 & 0.968\\
DGP 3         & 0.833 & 0.833 & 0.654 & 0.854 & 0.646 & 0.847 & 0.857 & 0.563 \\
DGP 4         & 0.709 & 0.830 & 0.673 & 0.701 & 0.700 & 0.618 & 0.621 & 0.835 \\
DGP 4b        & 0.897 & 0.856 & 0.673 & 0.701 & 0.716 & 0.618 & 0.621 & 0.835\\
   \hline
\end{tabular}
\end{table}

\begin{figure}[htp]
\centering
\includegraphics[width=0.48\textwidth]{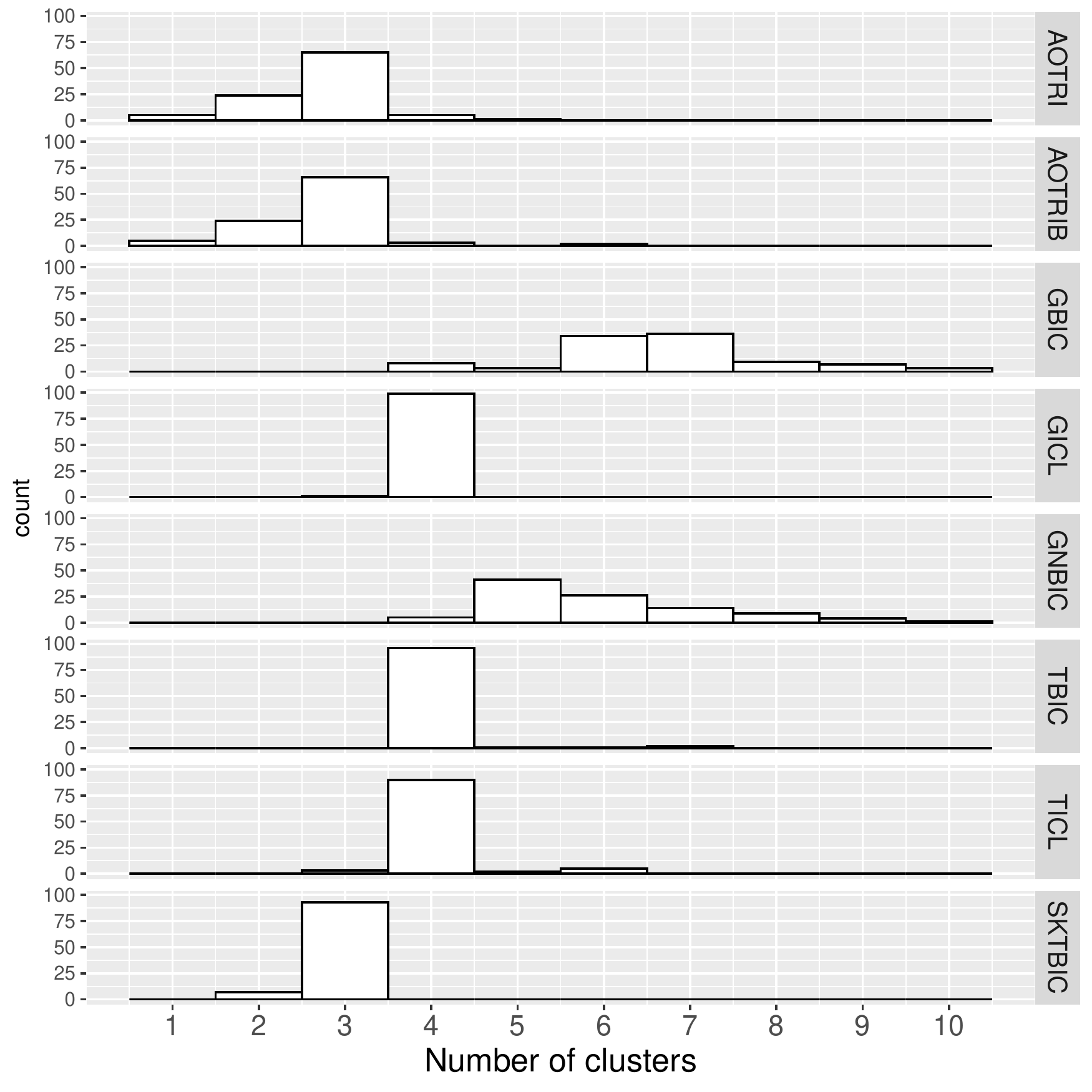}
\includegraphics[width=0.48\textwidth]{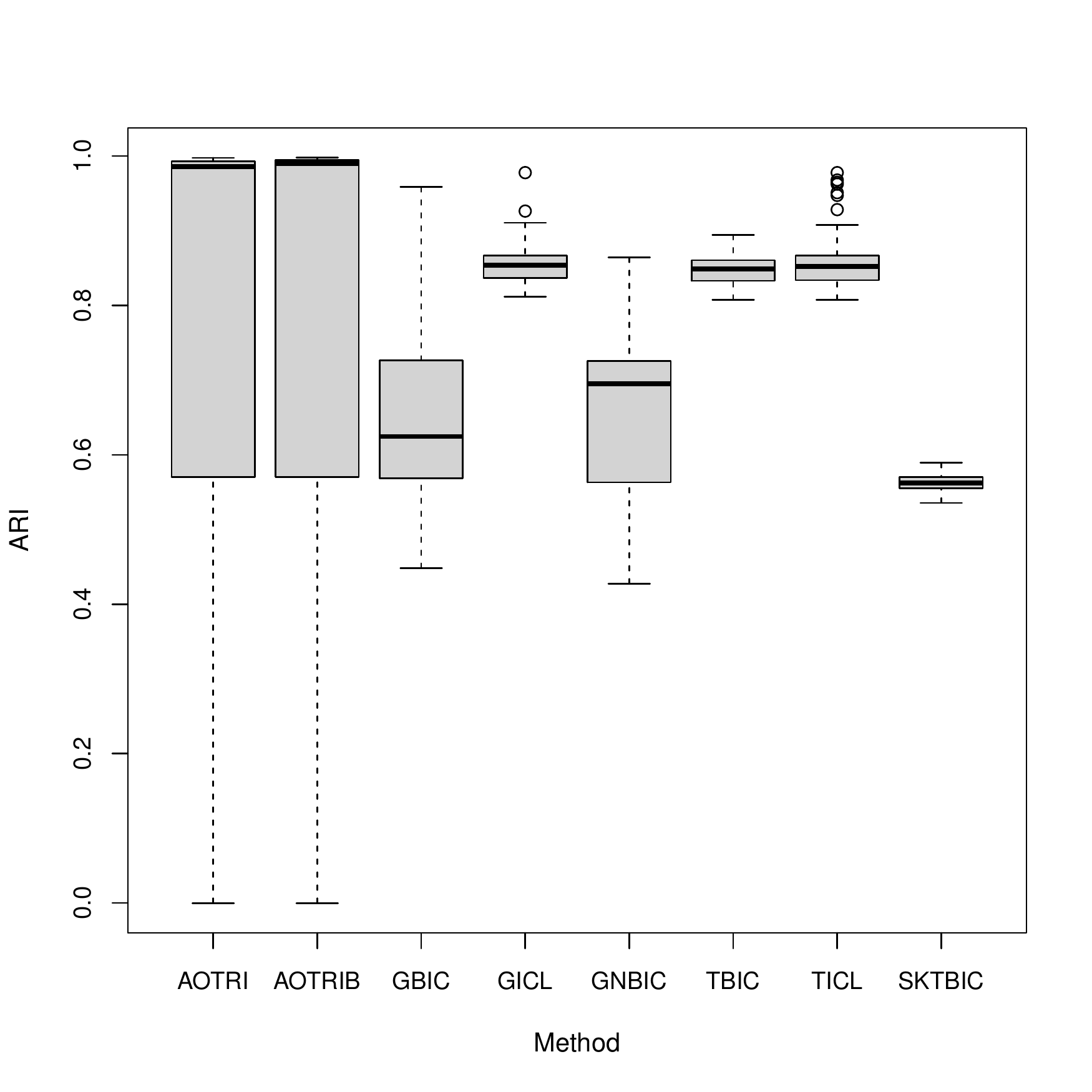}
\caption{\label{fig:tg3hresults} Left side: Distribution of numbers of
clusters by method for DGP 3 (true $G=3$) over 100 simulation runs. Right side:
Corresponding distribution of adjusted Rand index values. 
}
\end{figure}

DGP 3 with $n=2000, p=20$
was designed to deviate from the model assumptions in a way that does not
make the clusters look strikingly different from Gaussian ones, but with some
heavier tails.  Again the clustering structure is present 
only in the first two variables, but these are now $t_3$-distributed; variable
3-20 are again standard Gaussian; outliers as occasionally generated by 
$t_3$-distributions are now in the same variables that also have the clustering
structure, as opposed to DGP 1 and 2. See the supplement of \cite{CorHen16}
for full details. Figure \ref{fig:t3mclust} shows the first
two variables generated by this DGP. Results are shown in Figure 
\ref{fig:tg3hresults} and Table \ref{tab:arimean}. 

The AOTRI versions estimate $G=3$ correctly
for 69 data sets, and get the clustering almost completely right in these cases,
which does not hold for any of the other methods. {\color{black}
In a few situations they 
deliver a one-cluster solution with ARI$=0$ or a two-cluster solution, 
which makes the overall ARI mean worse than that of GICL and the t-mixture 
based methods. As this DGP has t- as well as Gaussian distributions, the model
assumptions of none of these methods is perfectly fulfilled. 
They estimate $G=4$ in almost all cases, adding a fourth mixture
component that collects observations so that the three main components
look closer to the assumptions. GBIC and GNBIC overestimate 
$G$ more strongly, and
yield clearly worse ARI results, as does SKTBIC, despite correctly
estimating $G=3$ in most cases.}

\begin{figure}[htp]
\centering
\includegraphics[width=0.8\textwidth]{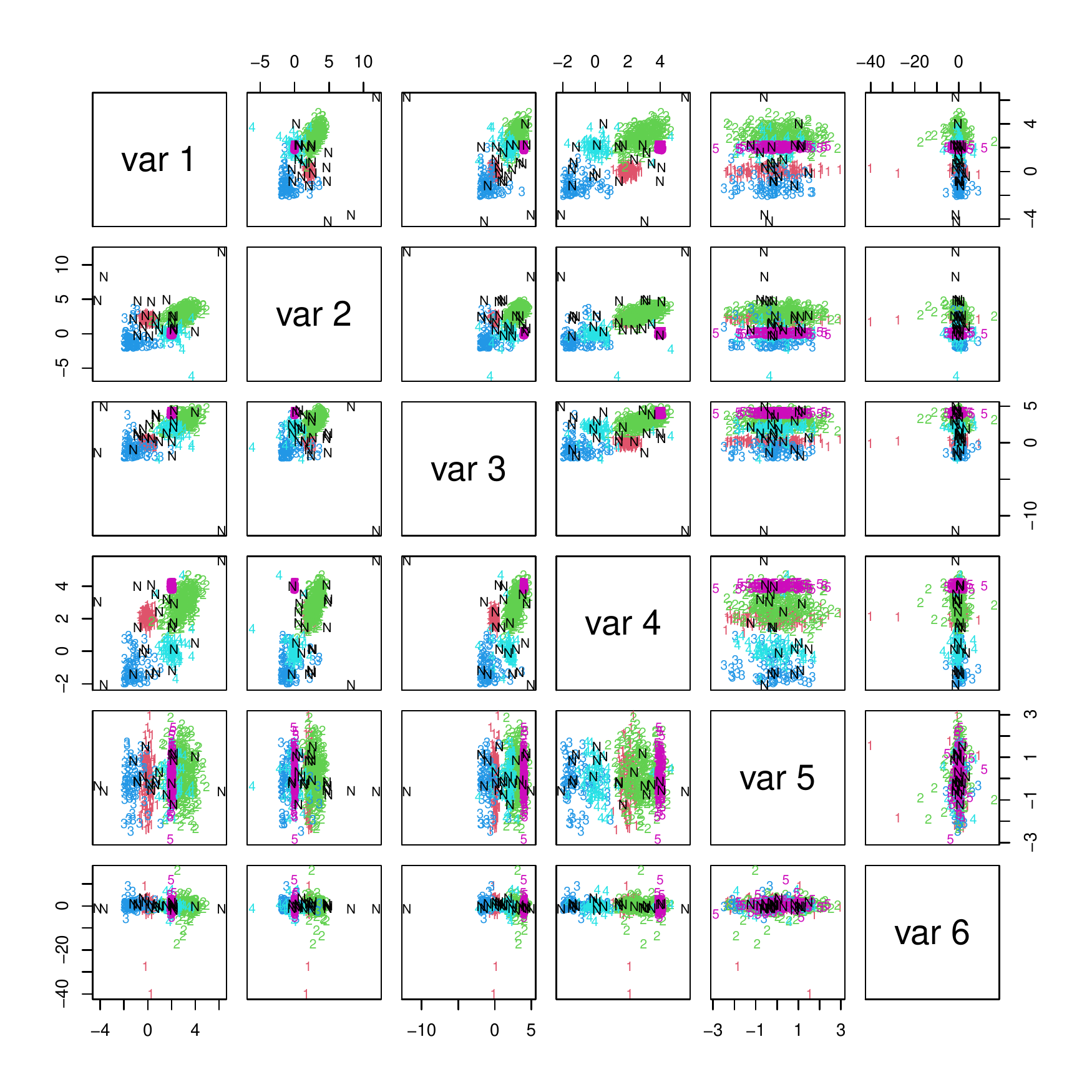}
\caption{\label{fig:dgp3data} Data simulated from DGP 4 with true clustering. ``N'' denotes noise, half of which was generated by a uniform and half by a $t_3$, see \cite{Hennig07}.}
\end{figure}

DGP 4 with $n=660, p=6$ brings together different shapes of distributions in the same data set, as is the case in some real applications. Cluster structure occurs on the first four variables, the fifth variable is standard Gaussian, the 
sixth is $t_2$, generating some
outliers. There are two Gaussian clusters with sizes 250 and 150, an 
independent product of exponential variables with 70 observations, a shifted
multivariate $t_2$-distribution with 70 observations, and a tight uniform with
100 observations, therefore $G=5$. 
There are 20 ``true'' noise points, 10 of which are 
generated by a wide uniform distribution and 10 by a wider spread $t_2$, see 
Figure \ref{fig:dgp3data}. 
This was taken from \cite{Hennig07}, where details are given. Only the uniform
cluster was added, centered at $(2,0,4,4)$ with range 0.4 on the first 
four variables. 

\begin{figure}[htp]
\centering
\includegraphics[width=0.48\textwidth]{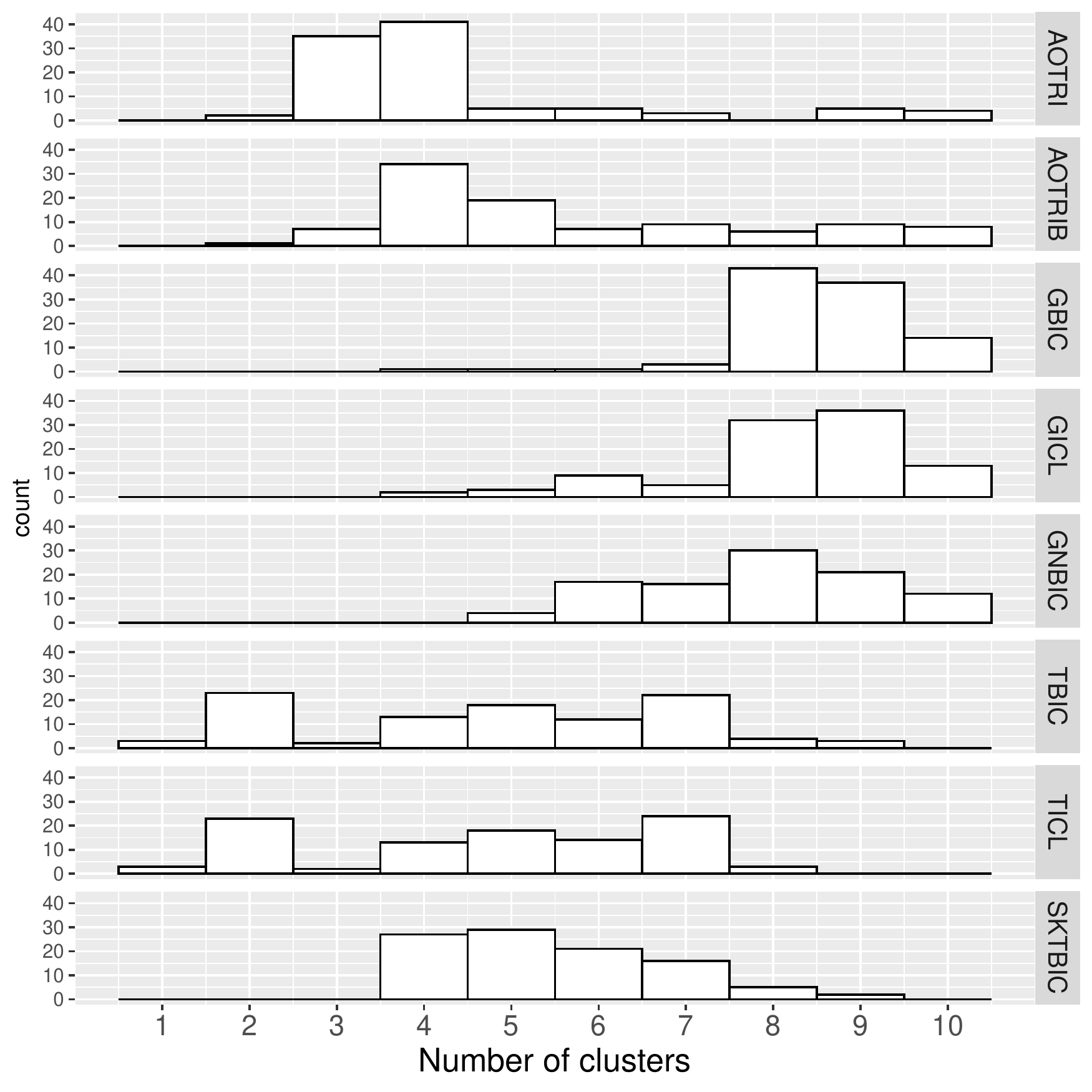}
\includegraphics[width=0.48\textwidth]{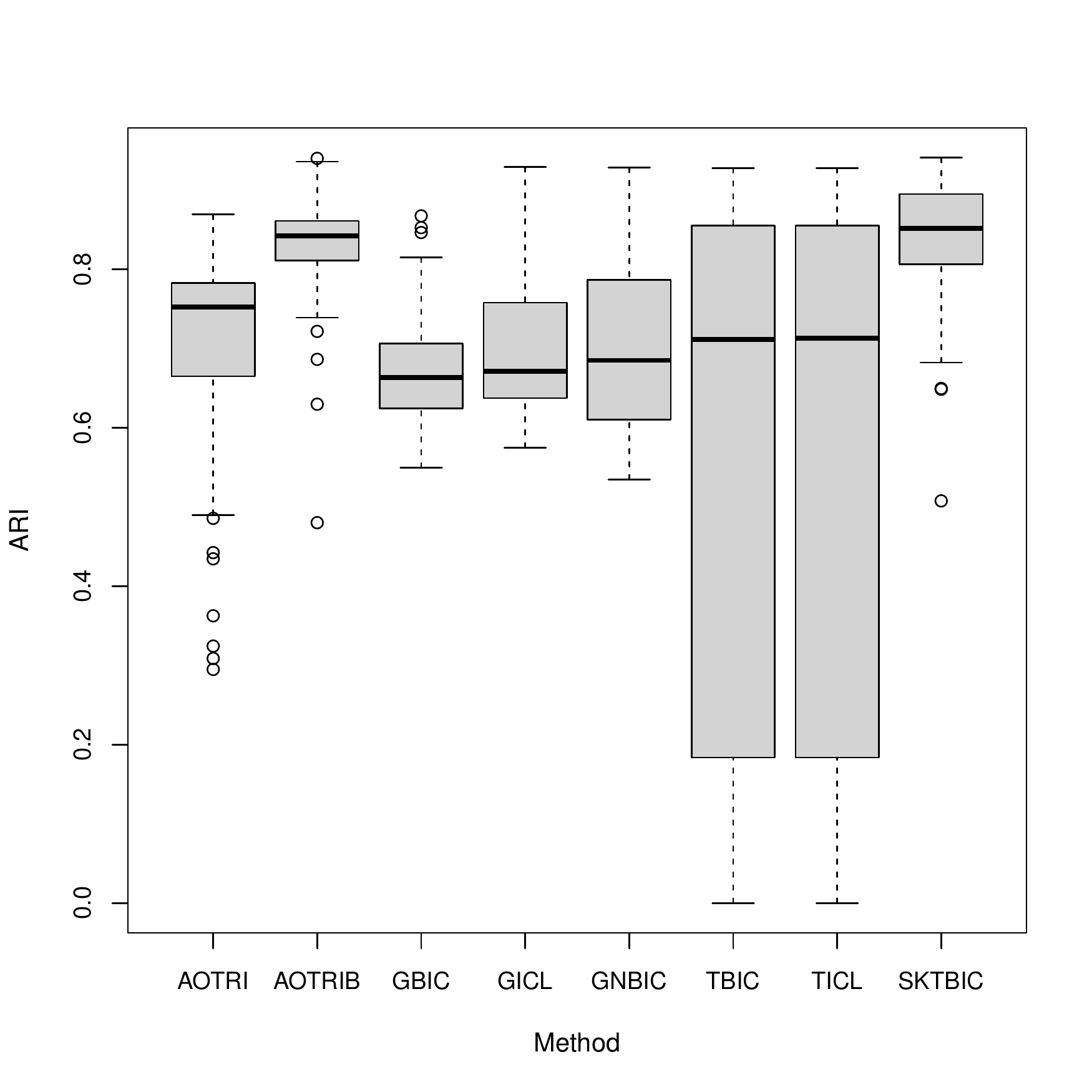}
\caption{\label{fig:test4number} Left side: Distribution of numbers of
clusters by method for DGP 4 (true $G=5$) over 100 simulation runs. Right side:
Corresponding distribution of adjusted Rand index values. 
}
\end{figure}

{\color{black}
For the results see Figure \ref{fig:test4number} and Table \ref{tab:arimean}.
The best performance is shown by SKTBIC regarding choosing $G$ and also 
regarding 
the plain ARI. AOTRI and AOTRIB have a tendency to underestimate the
number of clusters. This can mainly be explained by the fact that the strongly 
asymmetric exponential cluster is not 
well represented by a mode at the mean, and
therefore the $Q$-criterion will prefer solutions that classify this as noise.
This is not a proper cluster in the sense defined by $Q$ (as long as its
asymmetric 
version is used) and should arguably not be counted when 
operating with a symmetric prototype idea of a cluster. We also give ARI-results
not involving the observations classified as noise in Table \ref{tab:arimean}
(DGP 4b). Regarding these, AOTRI and 
AOTRIB perform better than the SKTBIC; if in a real 
application it is acceptable to not classify and interpret the observations
classified as ``noise'', AOTRI classifies the remaining observations very 
reliably. Regarding the number of clusters and raw ARI, AOTRIB 
with $\beta=1/3$ is almost as good as SKTBIC and clearly better than AOTRI. 
The latter is better when 
estimated noise is discounted. This is largely due to the larger 
estimated noise proportion: Observations that are not identified as noise by
AOTRI are those that are easier to classify. GBIC, GICL, and GNBIC tend 
to fit some non-Gaussian clusters
with more than one Gaussian component, and overestimate $G$ in this way. The
ICL does not help much here. TBIC and TICL produce a large variance of the ARI 
and the estimated $G$, sometimes 
over- and sometimes underestimating it.

Overall AOTRI and AOTRIB 
show a good and sometimes the best performance,
although they run into occasional problems (particularly when estimating $G=1$
in DGP 3). The remaining methods all have some strength and some weaknesses;
SKTBIC does very well except in DGP 3. GBIC and GNBIC overestimate $G$ in
case of non-Gaussian clusters, with deteriorating effect on the ARI. GICL
does somewhat better, but does not solve the issue completely. TBIC and TICL
do rather well for DGPs 2 and 3 but are much weaker in DGPs 1 and 4.}
 
\subsection{Olive oil data}\label{sec:srna}
{\color{black} 
The first real data set is from \cite{FoArLaTi83}. The data set contains 
$p=8$ chemical measurements on $n=572$ different specimen of olive oil 
produced in $G=9$ regions in Italy (northern Apulia, southern Apulia, 
Calabria, Sicily, inland Sardinia and coast Sardinia, eastern and western 
Liguria, Umbria). It has been used several times for benchmarking of supervised
and unsupervised classification methods. Interpreting the regions as the true
clusters, some of them have clearly non-Gaussian shape, and there are some 
outliers. Figure \ref{fig:olive18} shows two of the chemical variables,
``palmitic'' and ``eicosenoic'', where some peculiarities (an outlier
in region 7 on the lower left side of the plot, skewness or outliers
of region 4, a small and less dense subgroup on the right side of region 3)
can be seen.

\begin{figure}[htp]
\centering
\includegraphics[width=0.48\textwidth]{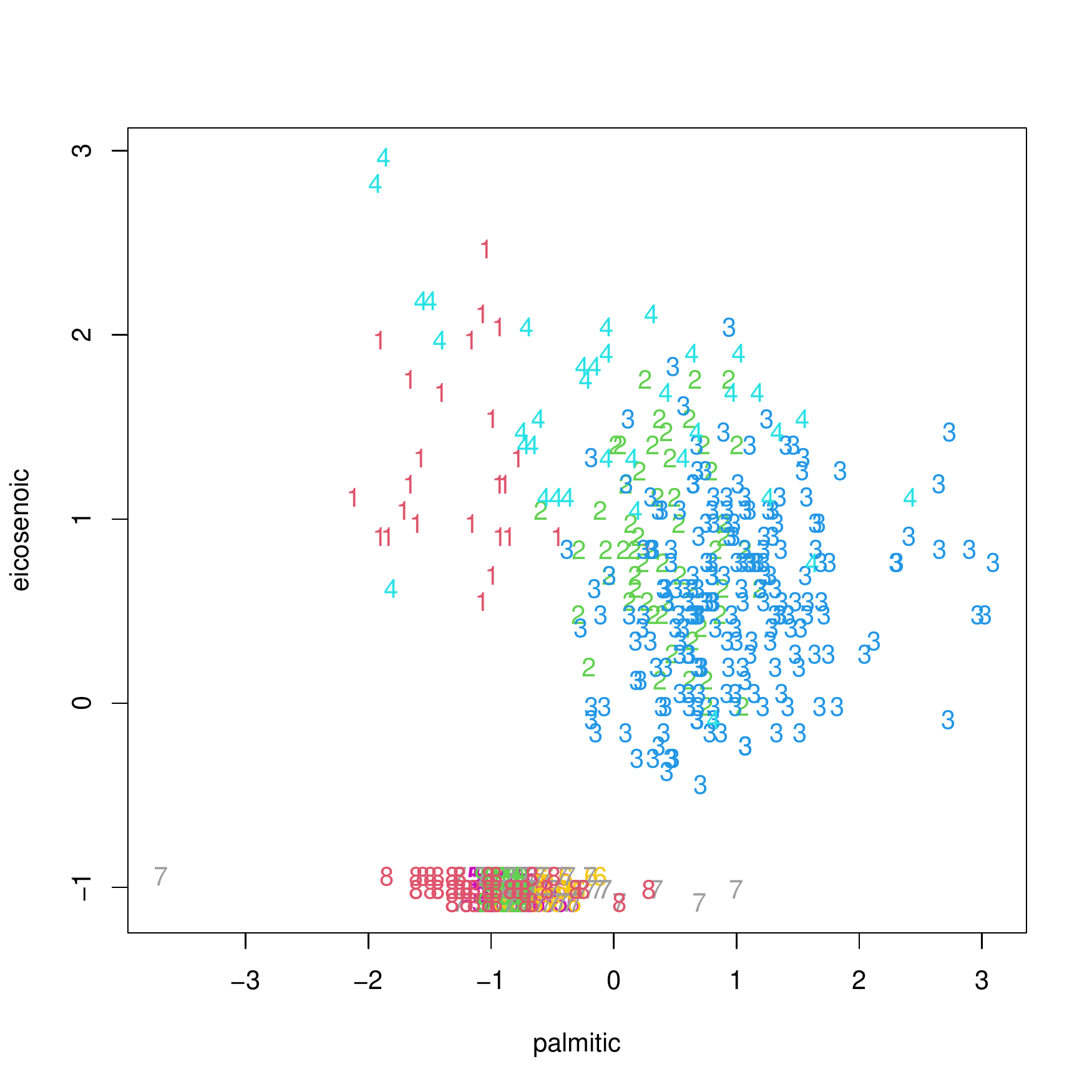}
\includegraphics[width=0.48\textwidth]{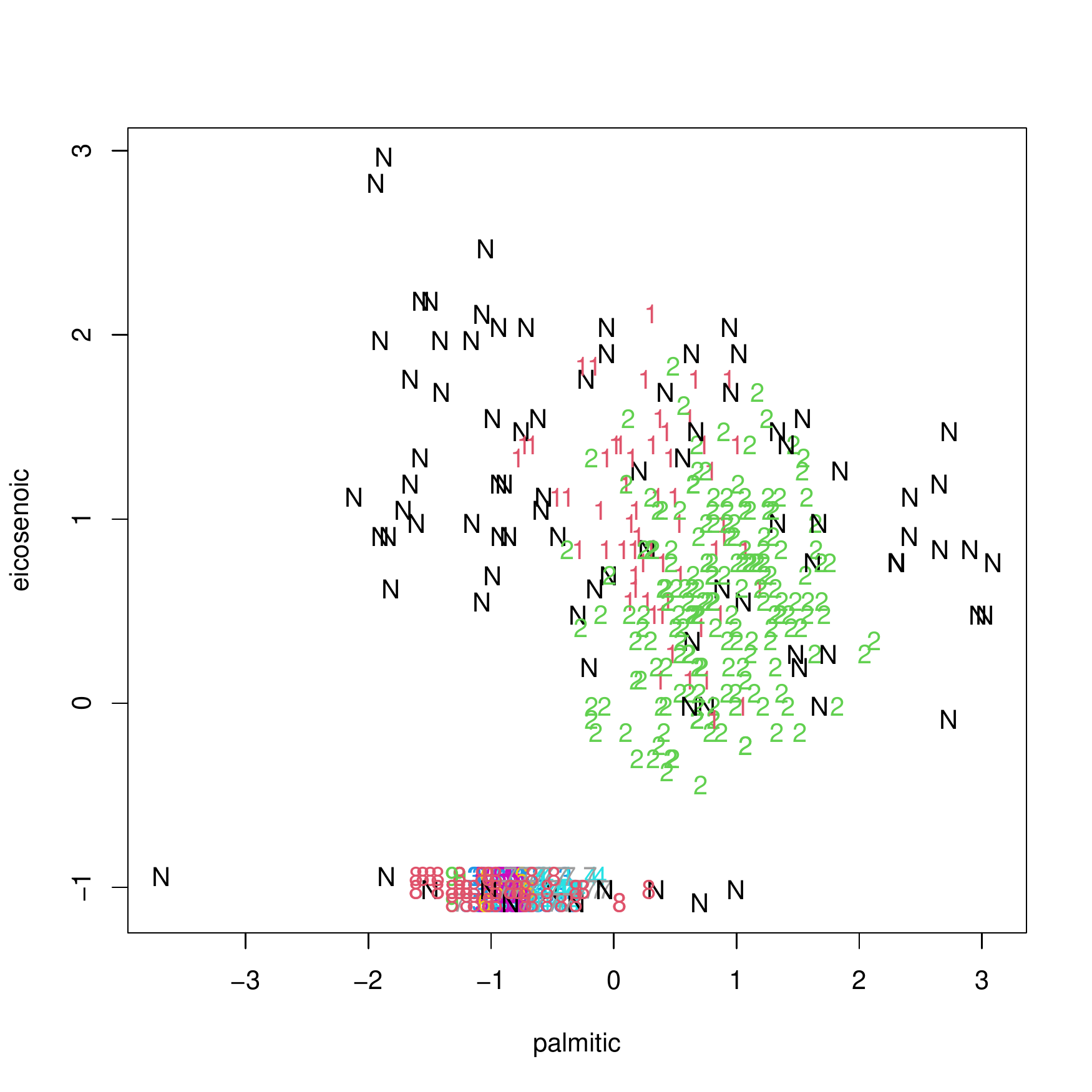}
\caption{\label{fig:olive18} Scatterplot of two of the eight variables in the
Olive Oil data set. Left side: true regions, right side: classification
by AOTRI.} 
\end{figure}

Assuming that a researcher analysing these data knows that clusters can be 
rather small (the smallest true cluster contains 4.4\% of the
observations) we analyse this using $p_0=0.02$. The AOTRI results are
visualised in Figure \ref{fig:oliveotrimle}. This shows that $G=3$ and $G\ge 6$
are adequate. $G=9$ has a substantially lower noise proportion than the lower
numbers of clusters. It is therefore the $G$ with the smallest $S(G)$ out of
the adequate ones, and is chosen as optimal, so that $G$ is estimated 
correctly. It still classifies more than 22\% 
of the observations as noise. 
The right side of Figure \ref{fig:olive18} shows
its classification for the variables ``palmitic'' and ``eicosenoic''. It can 
be seen that the earlier mentioned lower density subset of region 3 and the
visible outliers of region 7 are 
classified as noise, along with several olive oils from region 4 and 1 in a 
lower density region, and some observations that are outlying in other directions than those defined by these 
two variables. Data analytically this makes some sense. 22\% noise 
seems high, but other methods have issues with correctly classifying 
these observations as well. 
The ARI between this solution and the true regions is 0.762, including
the observations classified as noise; without them it is 0.891. AOTRIB
estimates $G=8$ with only 5.6\% of the observations classified as noise, and
it achieves a better ARI of 0.808 including the noise. Choosing $p_0=0.05$ is 
not advisable for these data. For
AOTRI this leads to an estimate of $G=6$ with an ARI 0f 0.712
as more noise is then tolerated, 
but it does not change the good result of AOTRIB, 
because the other numbers of clusters are inadequate with too high values of 
$Q(G)$.

\begin{figure}[htp]
\centering
\includegraphics[width=0.48\textwidth]{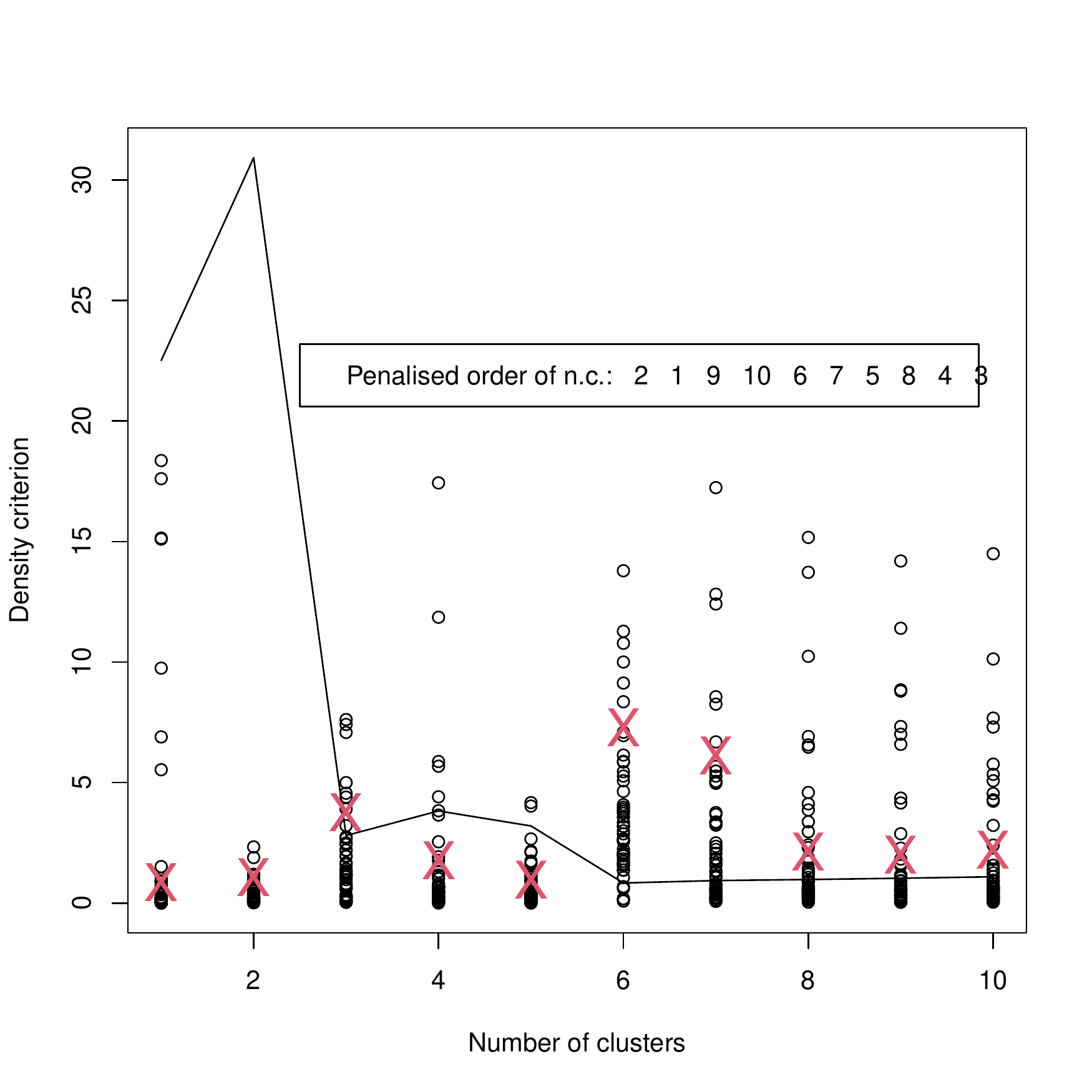}
\includegraphics[width=0.48\textwidth]{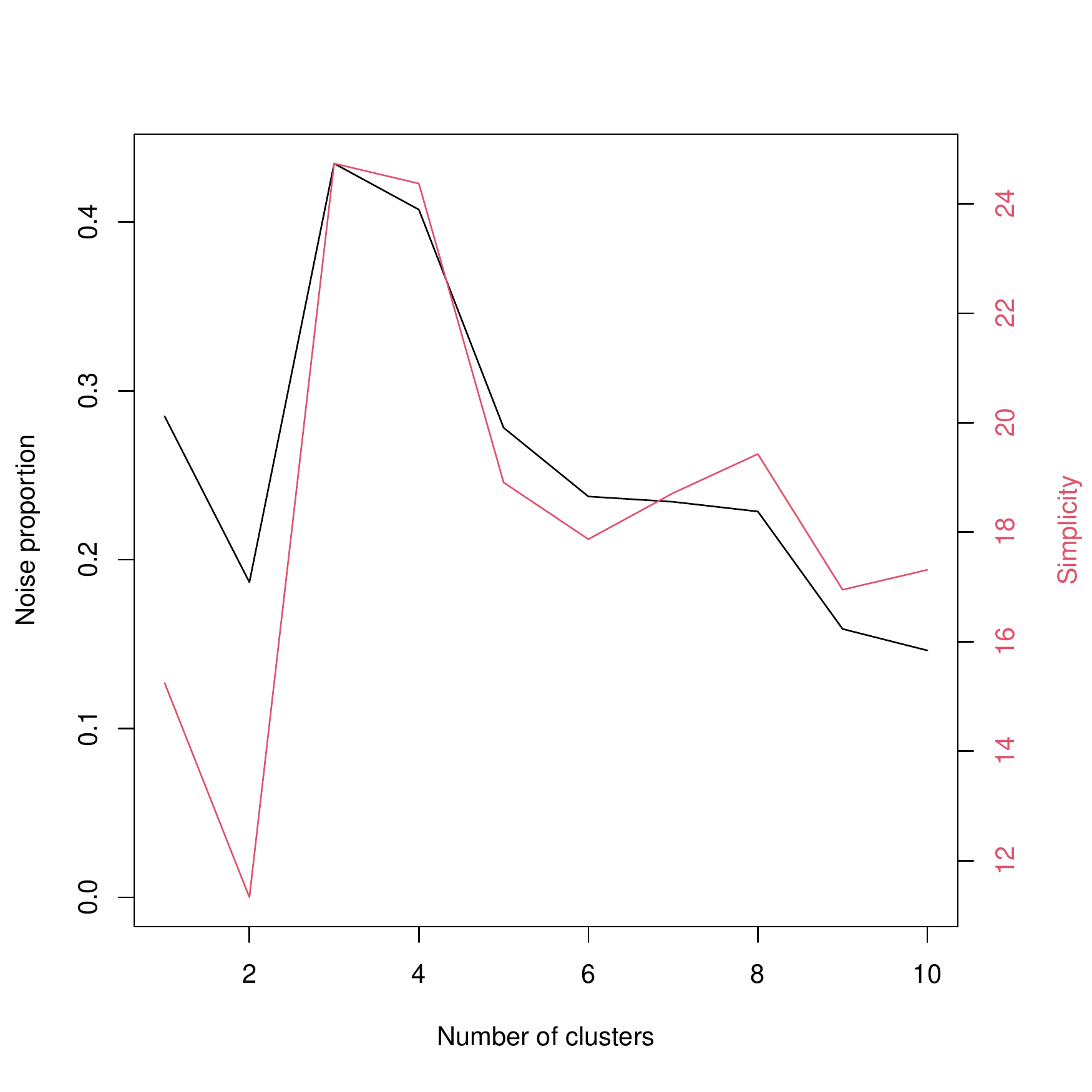}
\caption{\label{fig:oliveotrimle} AOTRI results for olive oil data. 
Left side: Density-based clustering quality
criterion $Q(G)$ for the different numbers of clusters. The 
connected lines refer
to the clustering of the original data set, the circles to the clustering
of bootstrapped data sets. The red ``X'' denotes the cutoff point for a clustering to be adequate. Right side:
Noise proportions, and ordering of numbers of clusters according to $S(G)$.
}
\end{figure}

GBIC and GICL both estimate the same model with $G=10$ and an ARI of 0.552 at
the upper end of the range of tried out $G$-values.
Introducing a noise component as GNBIC does improves this to 0.599 with $G=8$
including the noise, or 0.607 without it.
TBIC and TICL agree on $G=5$. The ARI is rather good at 0.773, although
not at the level of AOTRIB. SKTBIC estimates $G=7$ and an ARI of 0.548.}

\subsection{Districts of the city of Dortmund}\label{sec:simdata}
{\color{black}
A data set characterising the 170 districts of the German city of Dortmund is
presented in \cite{SomWei05}. This data set does not come with true cluster
labels. We used a version consisting of five sociological key variables 
and transformed them in such a way that fitting Gaussian distributions 
within clusters makes sense. The resulting variables are the logarithm of the unemployment rate (``unemployment''), the birth/death balance divided by number of inhabitants (``birth.death''), the migration balance divided by number of inhabitants (``moves.in.out''), the logarithm of the rate of employees paying social insurance (``soc.ins.emp''), and the percentage of foreigners among all 
unemployed and dependently employed persons (``foreigners'').

\begin{figure}[htp]
\centering
\includegraphics[width=0.8\textwidth]{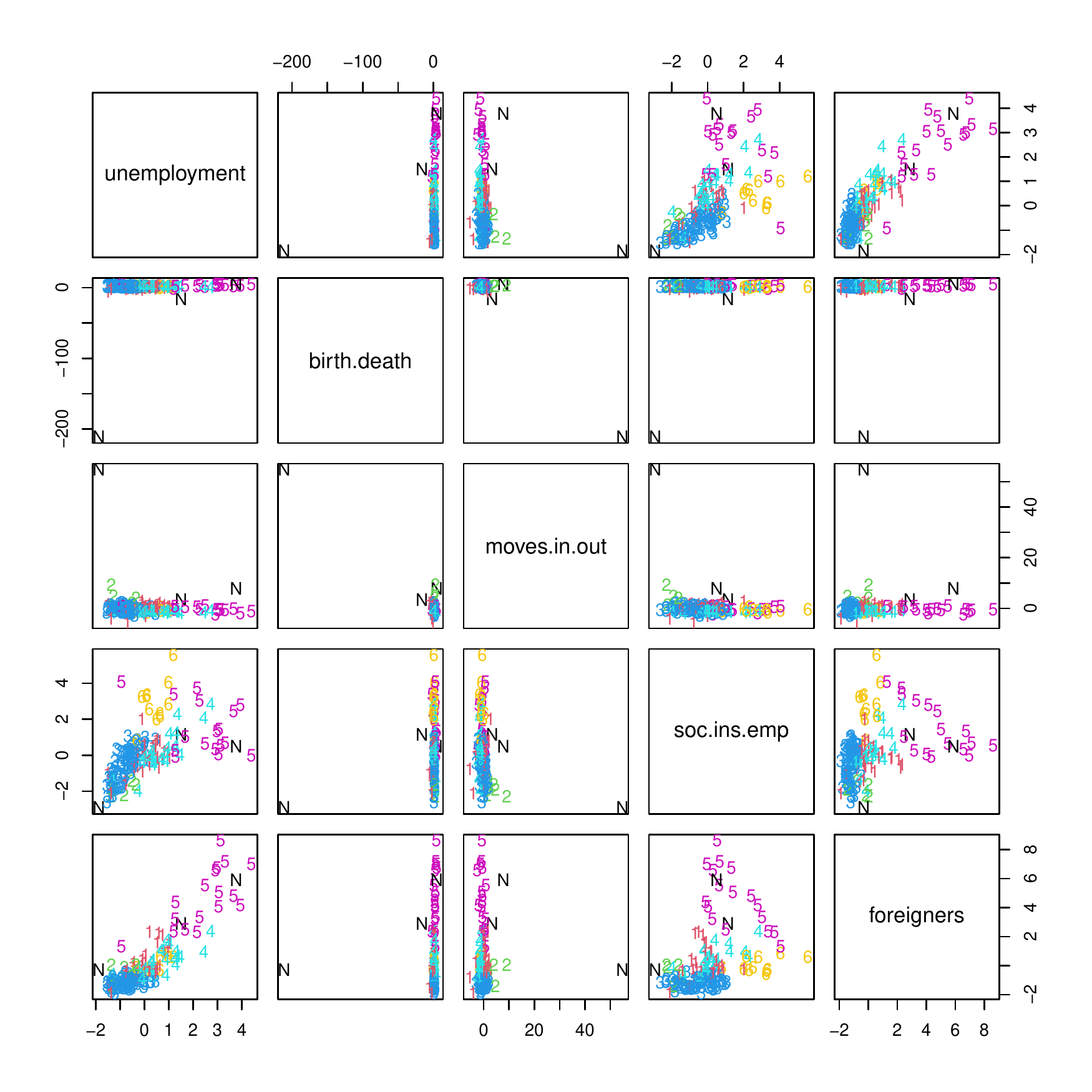}
\caption{\label{fig:dortmundpairs} District of Dortmund data with clustering
by AOTRI.}
\end{figure}

Figure \ref{fig:dortmundpairs} shows that there is an extreme outlier in the 
scatterplot of ``birth.death'' and ``moves.in.out'' (called ``Romberg Park'', 
a park district
with a clinic), with some more outlier 
candidates. The scatterplots of ``soc.ins.emp'' with ``unemployment'' and
``foreigners'', respectively, show some potential non-homogeneous structure.

\begin{figure}[htp]
\centering
\includegraphics[width=0.48\textwidth]{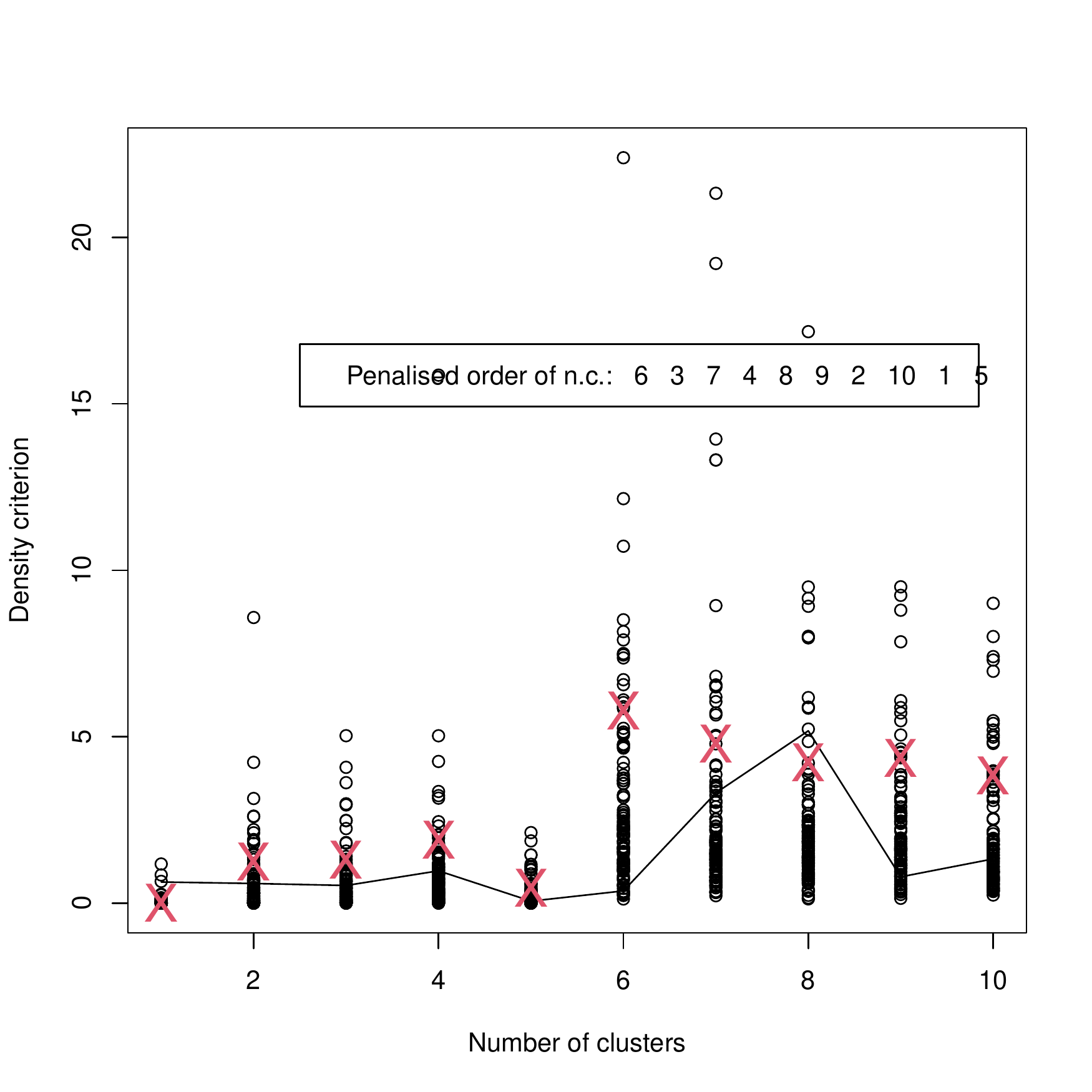}
\includegraphics[width=0.48\textwidth]{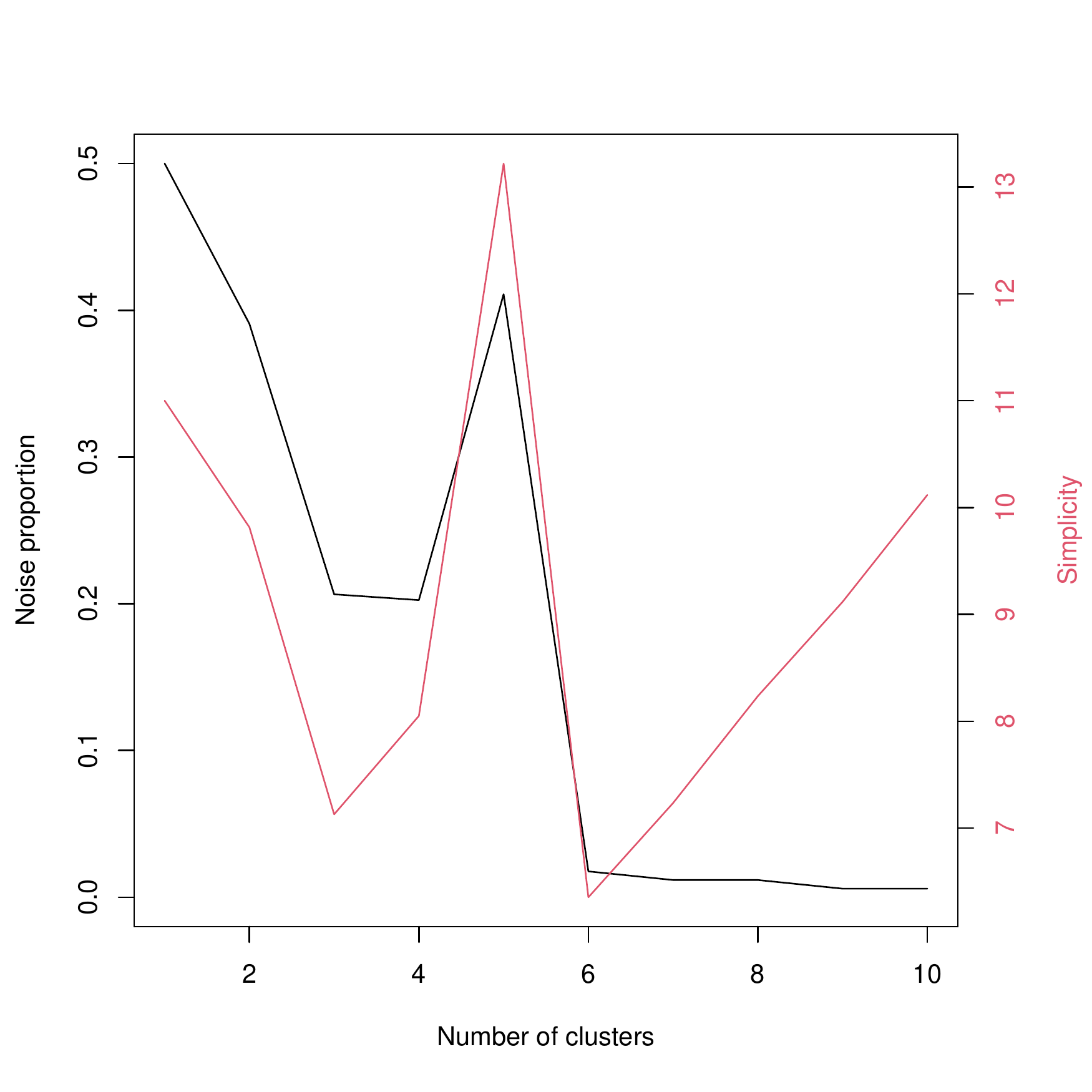}
\caption{\label{fig:dortmundotrimle} AOTRI results for Dortmund data. 
Left side: Density-based clustering quality
criterion $Q(G)$ for the different numbers of clusters. The 
connected lines refer
to the clustering of the original data set, the circles to the clustering
of bootstrapped data sets. The red ``X'' denotes the cutoff point for a clustering to be adequate. Right side:
Noise proportions, and ordering of numbers of clusters according to $S(G)$.
}
\end{figure}

Figure \ref{fig:dortmundotrimle} shows that the clusterings for 
$G$ between 2 and 7, 9, and 10 are assessed as adequate, but that there are
very high noise proportions for $G<6$. Therefore AOTRI assesses $G=6$ as
optimal. As far as the clusters can be assessed from Figure 
\ref{fig:dortmundpairs}, they seems sensible, with clusters 5 and 6 
showing a particularly clear profile, although one may wonder whether
the data could be represented by a lower number of clusters. The outliers 
seem well detected.
Figure \ref{fig:dortmundbox} shows that the different clusters have distinct
profiles in terms of the five variables. Cluster 5 has the highest values
of ``unemployment'', ``foreigners'', and ``birth.death''. Cluster 6 has high
``social.ins.emp'' and is quite homogeneous (though not extreme) on a number of
other variables. Cluster 2 is highest on ``moves.in.out'' and lowest on 
``social.ins.emp''. Cluster 3 is lowest on ``foreigners'' and joint lowest on 
``unemployment''. Cluster 4 is lowest on ``moves.in.out'' and has otherwise
fairly homogeneous values in the middle of the range. Cluster 1 is more 
difficult to interpret, with homogeneous mid-range values for ``social.ins.emp'' and ``foreigners'' and a large variance including the lowest values of
``birth.death''.

\begin{figure}[htp]
\centering
\includegraphics[width=0.32\textwidth]{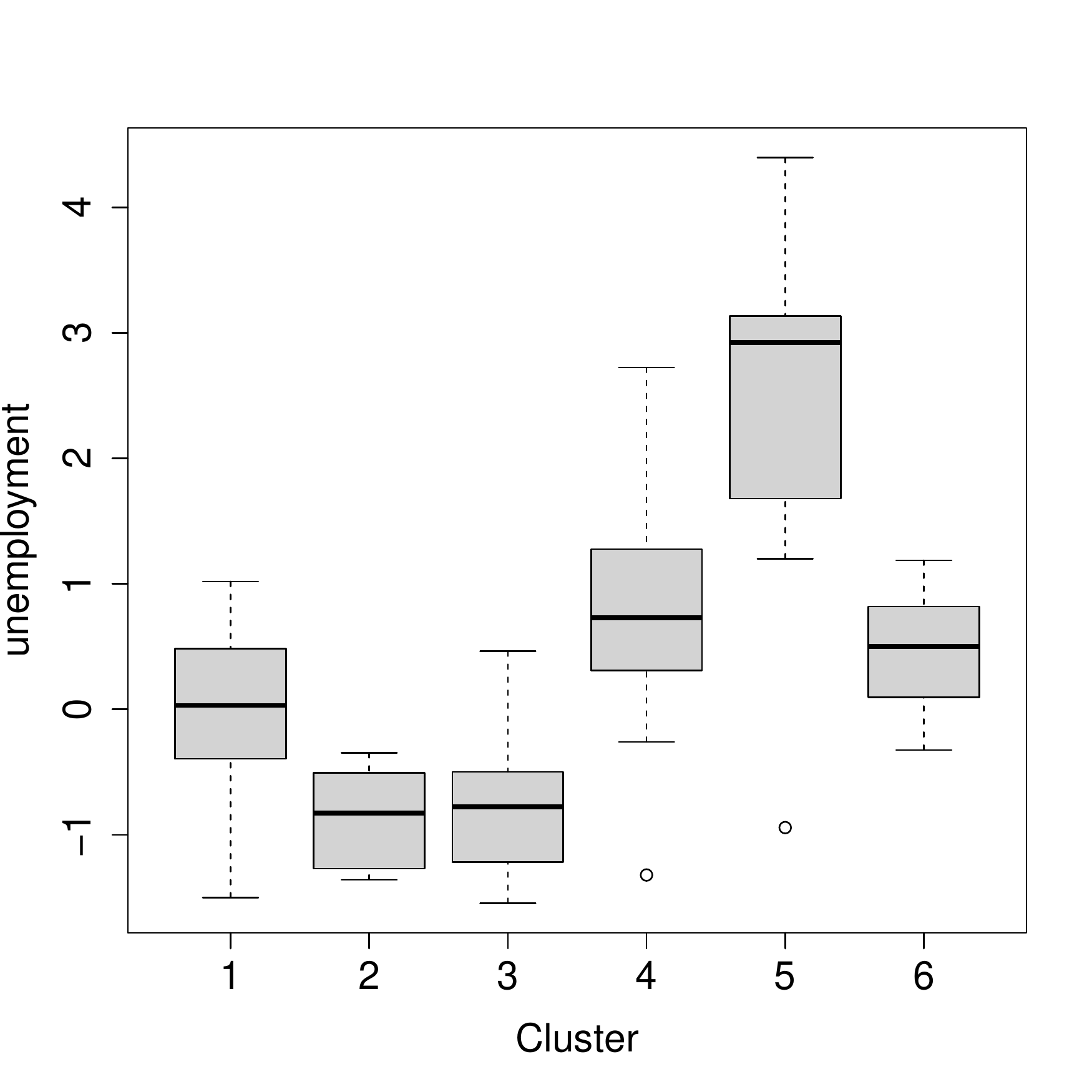}
\includegraphics[width=0.32\textwidth]{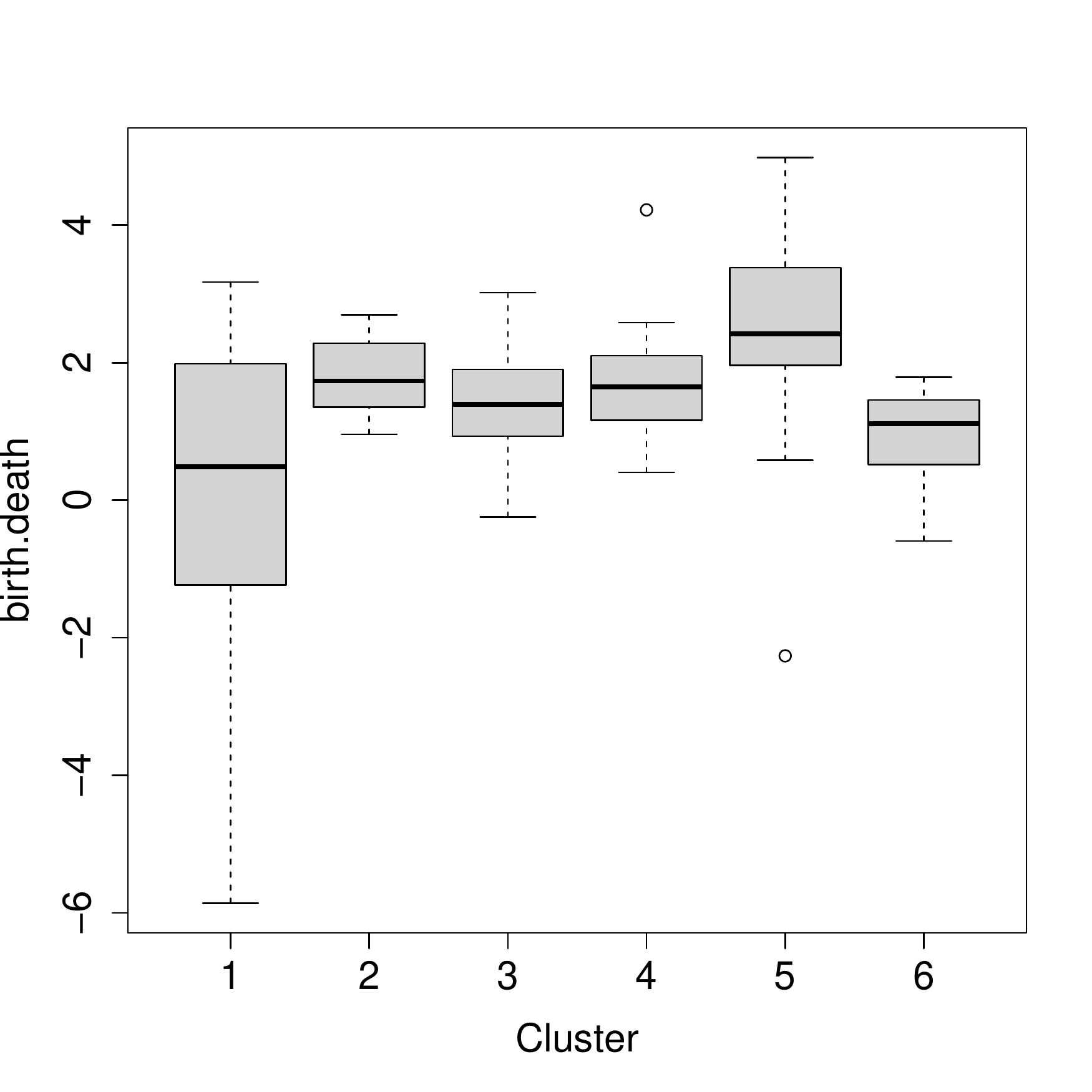}
\includegraphics[width=0.32\textwidth]{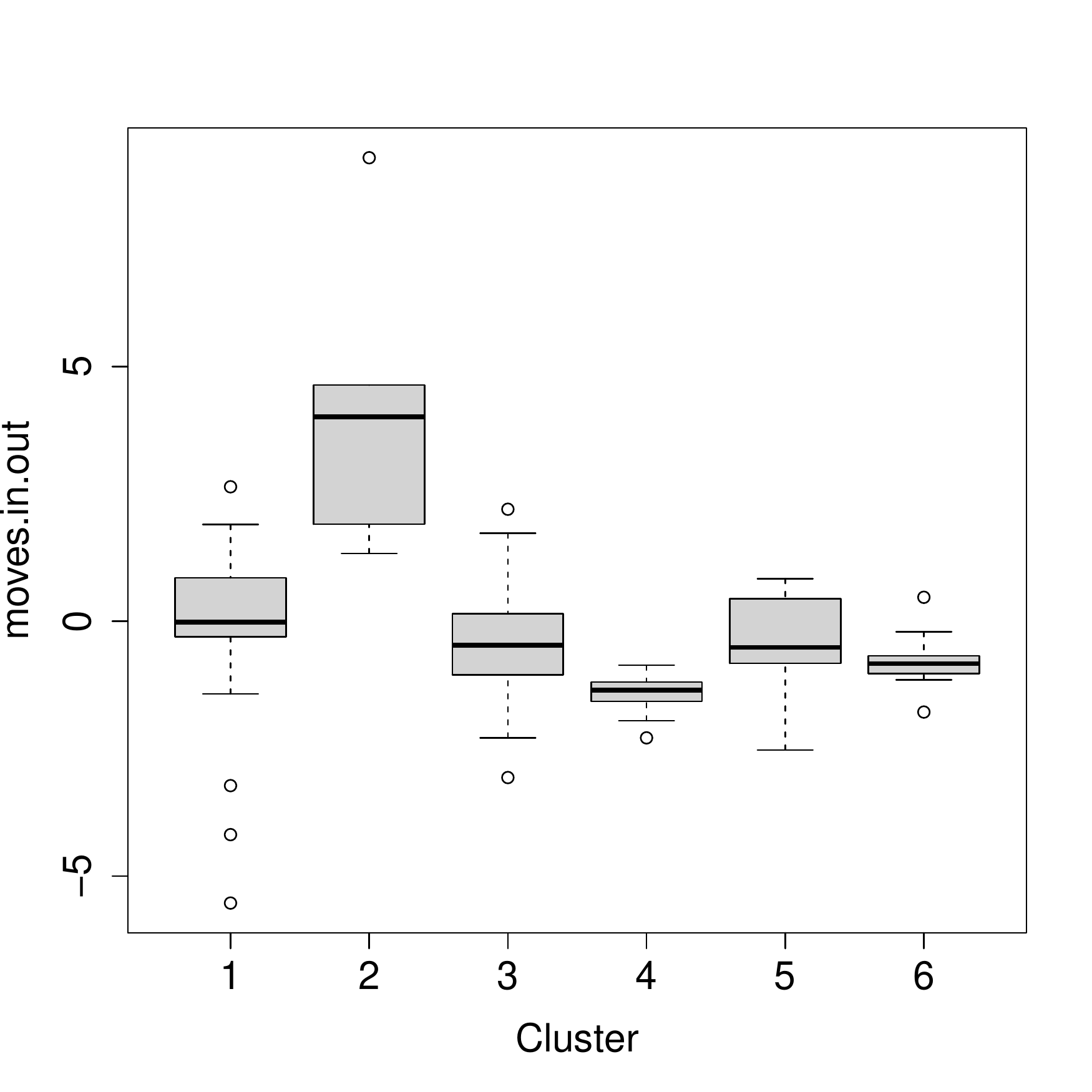}
\includegraphics[width=0.32\textwidth]{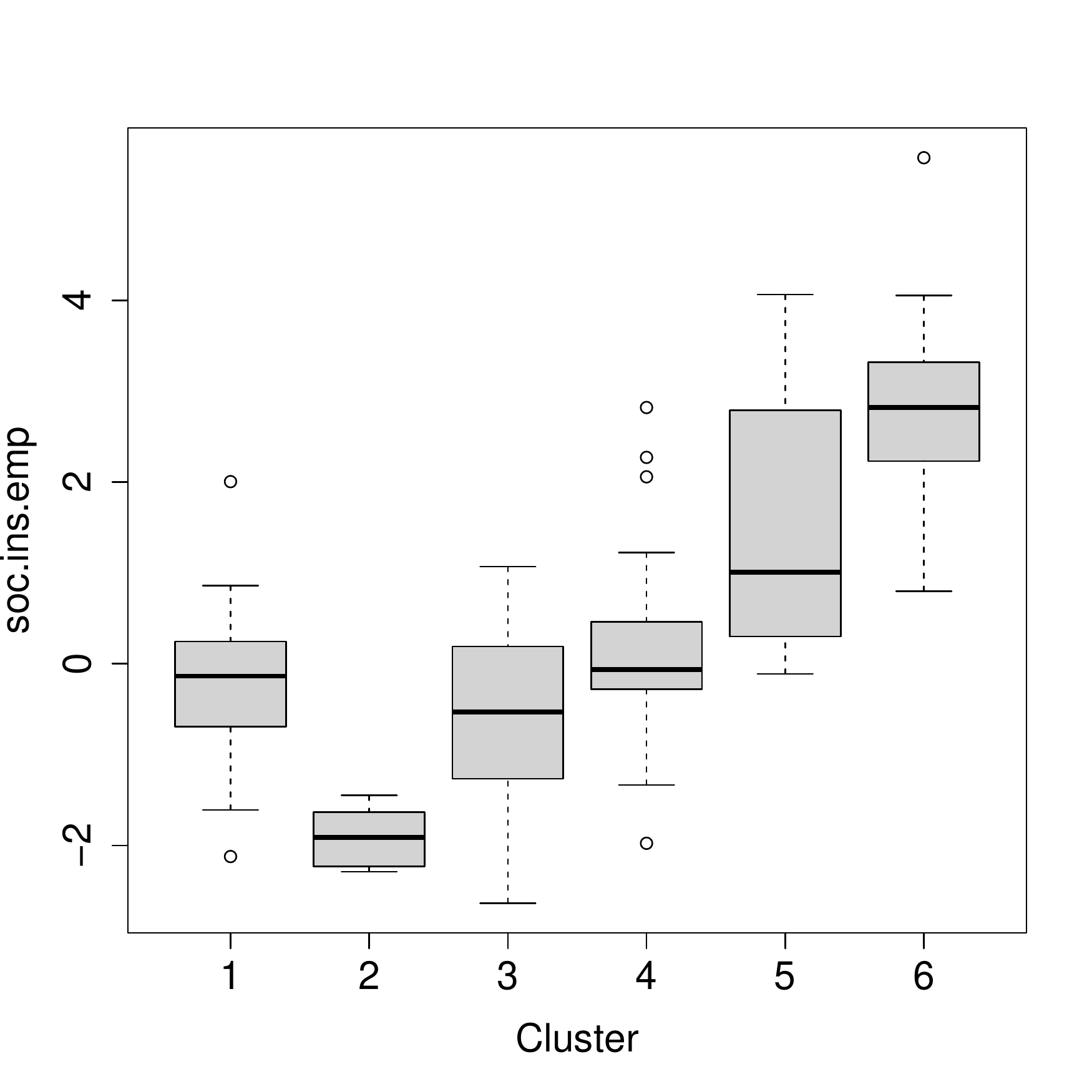}
\includegraphics[width=0.32\textwidth]{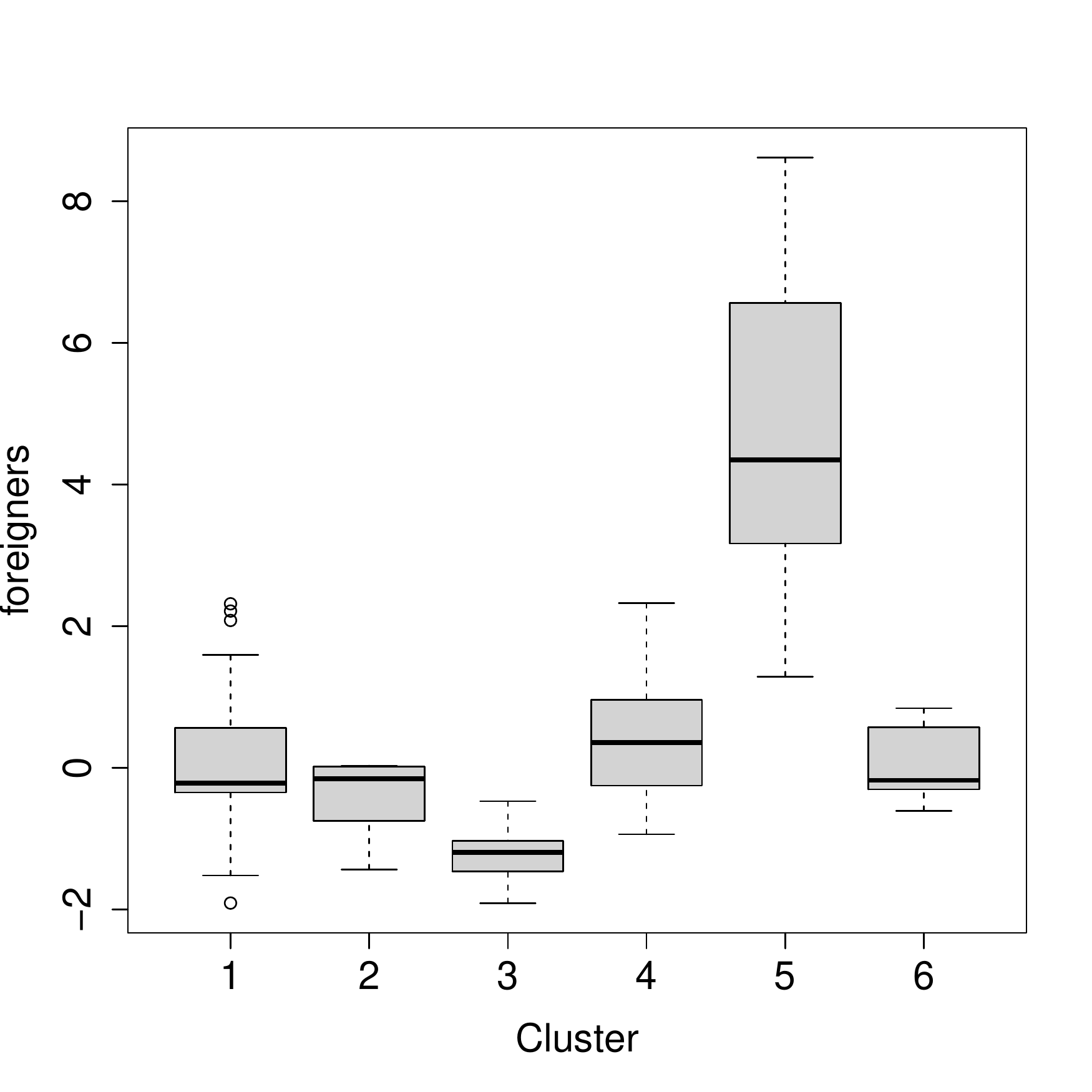}
\caption{\label{fig:dortmundbox} Boxplots of the five variables of the 
Dortmund data grouped by the 6 clusters from AOTRI.
}
\end{figure}

AOTRIB finds the same outliers as AOTRI, but prefers $G=3$ (with noise 
proportion 0.018, whereas AOTRI estimates this as 0.206 for $G=3$), putting the 
observations in the lower
density region in the upper right of ``unemployment'' vs. ``soc.ins.emp'',
i.e. AOTRI's clusters 5 and 6 and parts of cluster 4, together, and splits
the rest into two bigger clusters, one of which is almost identical to 
AOTRI's cluster 3. This seems data analytically reasonable with the clusters
more visibly 
distinct, although it encodes
rougher information on the structure of the districts.
GBIC and GICL choose the same model with $G=4$ as optimal. This includes a 
large variance cluster joining the 
outlier Romberg Park with some districts that have high ``foreigners'' values
and does not make much sense. TBIC and TICL choose $G=2$, just distinguishing
the main bulk of the data from a group collecting atypical observations in
various directions. SKTBIC is flexible enough to fit
the whole data set by $G=1$. All of these seem to be of little use for the
understanding of the city districts. GNBIC with noise components 
selects a reasonable solution with $G=4$, compromising between AOTRI and 
AOTRIB, finding one outlier more. 

Overall the simulations and data examples do not highlight AOTRI/AOTRIB as 
universally best method, but rather illustrate that it can give reasonable and 
useful results in a variety of situations in which several competitors have
difficulties.}
\section{Conclusion}\label{sec:conclusion}
The problem of choosing the number of 
clusters is very difficult, particularly in applications in which observations 
occur that do not belong to any cluster. It is often treated as an estimation 
problem regarding the true number of mixture components in a parametric mixture
distribution, e.g., a Gaussian mixture, but then clusters that make 
interpretative sense and are even slightly non-Gaussian are often fitted by more
than one mixture component. 

An appropriate decision rule for the
number of clusters in a Gaussian mixture context involves a decision about 
what kind of non-Gaussian data subset still qualifies as a cluster. This is 
formalised by our clustering quality statistic $Q$. The observed value of $Q$ is
compared to what is expected if data are indeed generated by a Gaussian mixture
with the estimated parameter values. If an underlying distribution of a cluster
has a tendency to produce better clusters than a Gaussian according to $Q$ 
(which is the case for distributions such as the t-distribution, for which the 
density goes down faster from the mean than for the Gaussian), the
procedure will accept such clusters. Some users may be willing to accept certain
potentially unimodal clusters even though they look somewhat worse than what 
is expected 
from the Gaussian. This could be achieved by changing the cutoff value $c$ for
adequacy to something larger, say from 2 to 3 or 4. However this would allow for
clusters that look less unimodal. Another possible modification is to 
re-define $Q$ in order to allow for asymmetric clusters, although it may then
be better to start with a mixture of skew distributions. 

{\color{black} Readers may wonder whether the Gaussian mixture model is a 
good starting point if
it is of interest to fit non-Gaussian clusters by a single mixture component.
The answer is that this is appropriate if the interest is in finding clusters 
that are roughly Gaussian-shaped, which we define here as unimodal, and
approximately elliptical. We want to avoid modelling 
clusters that share enough key characteristics with the Gaussian, at least 
approximately, by more than one mixture component, which is the reason
why we do not choose $Q$ as a likelihood ratio or a goodness-of-fit statistic
for a Gaussian distribution.
Furthermore in many applications it is desirable to
have a distribution with light tails as a cluster prototype distribution, 
because distributions with heavier tails generate observations with larger
probability that are far from the main bulk of the data, and are therefore
often more appropriately interpreted as outliers rather than cluster members,
see \cite[p. 231 ff.]{McLPee00} for mixtures of t-distributions.}

It is ultimately up to the user to decide what kind of 
clusters are required in a given application. Without such decisions, the 
data on their own do not provide sufficient information about the clustering
structure required to fit them; there are severe identifiability problems when
choosing a mixture model. 

The general adequacy approach presented here
can be used for choosing the number of clusters for
other clustering methods, as long as a model is given that formalises 
a prototype clustering structure of interest to which parametric bootstrap 
can be applied. {\color{black} The clustering quality statistic $Q$ can be chosen
in different ways, formalising other concepts of admissible clusters,
or even the same concept in alternative ways. One could also attempt to 
select parameters such as $\delta$ and $\gamma$ in this way, although this
is probably more difficult due to the continuous nature of these parameters.
This is left to future work.}
 
The approach as presented here along with the accompanying plots shown in 
Sections 
\ref{sec:srna} and \ref{sec:simdata} will be implemented in the R-package 
\texttt{otrimle} by the time of the appearance of the accepted version of
this paper in the journal.

\bibliographystyle{chicago}


\end{document}